\documentclass[9pt,twocolumn,twoside]{pnas-new}

\templatetype{pnasresearcharticle} 
\setboolean{displaywatermark}{false}

\usepackage{amssymb,amsfonts,amsmath}
\usepackage[utf8]{inputenc}

\usepackage{subcaption}
\usepackage{bbm}
\usepackage{verbatim}
\newcommand{\beginsupplement}{%
        \setcounter{table}{0}
        \renewcommand{\thetable}{S\arabic{table}}%
        \setcounter{figure}{0}
        \renewcommand{\thefigure}{S\arabic{figure}}%
     }

\usepackage{braket}
\usepackage{bm}
\usepackage{mathrsfs}
\usepackage{color}
\usepackage{leftidx}
\usepackage{soul,xcolor}
\usepackage{comment}
\usepackage{multirow} 
\usepackage{cleveref}
\usepackage[normalem]{ulem}
\usepackage{placeins}
\usepackage[scr=boondox]{mathalfa}

\crefrangelabelformat{equation}{(#3#1#4--#5#2#6)}
\crefname{equation}{Eq.}{Eqs.}
\Crefname{equation}{Equation}{Equations}

\begin{document}

\title{Unveiling Hidden Orders: Magnetostriction as a Probe of Multipolar-Ordered States}

\author[a]{Adarsh S. Patri}
\author[b,c]{Akito Sakai}
\author[d]{SungBin Lee} 
\author[a]{Arun Paramekanti}
\author[b,c]{Satoru Nakatsuji}
\author[a]{Yong Baek Kim} 

\affil[a]{Department of Physics and Centre for Quantum Materials, University of Toronto, Toronto, Ontario M5S 1A7, Canada}
\affil[b]{Institute for Solid State Physics, University of Tokyo, Kashiwa, Chiba 277-8581, Japan}
\affil[c]{CREST, Japan Science and Technology Agency (JST),
4-1-8 Honcho Kawaguchi, Saitama 332-0012, Japan}
\affil[d]{Department of Physics, Korea Advanced Institute of Science and Technology, Daejeon, 34141, Korea}

\leadauthor{Patri} 



\correspondingauthor{\textsuperscript{1}To whom correspondence should be addressed. E-mail: ybkim@physics.utoronto.ca}


\begin{abstract}
Broken symmetries in solids involving higher order multipolar degrees of freedom are historically referred to as ``hidden orders'' due to the formidable task of detecting them with conventional probes. Examples of such hidden orders include spin-nematic order in quantum magnets, and quadrupolar or higher multipolar orders in various correlated quantum materials. In this work, \textcolor{black}{we theoretically propose that the study of magnetostriction provides a powerful and novel
tool to directly detect higher-order multipolar symmetry breaking} -- such as the elusive octupolar order -- by
examining its scaling behaviour with respect to an applied magnetic field $h$. 
As an illustrative example, we examine such key scaling signatures in the context of Pr-based cage compounds with strongly correlated $f$-electrons, Pr(Ti,V,Ir)$_2$(Al,Zn)$_{20}$, whose low energy degrees of freedom are composed of purely higher-order multipoles: quadrupoles $\mathcal{O}_{20,22}$ and octupole $\mathcal{T}_{xyz}$. Employing a symmetry-based Landau theory of multipolar moments coupled to lattice strain fields, we demonstrate that a magnetic field applied along the [111] direction results in a length change with a distinct linear-in-$h$ scaling behaviour, accompanied by hysteresis, below the octupolar ordering temperature. We show that the resulting ``magnetostriction coefficient'' is directly proportional to the octupolar 
order parameter, providing the first clear access to this subtle order parameter.
Along other field directions, we show that the field dependence of the magnetostriction provides a window into quadrupolar orders.
Our work provides a springboard for
future experimental and theoretical investigations of multipolar orders and their quantum phase transitions in a wide variety of systems.
\end{abstract}


\verticaladjustment{-2pt}

\maketitle
\ifthenelse{\boolean{shortarticle}}{\ifthenelse{\boolean{singlecolumn}}{\abscontentformatted}{\abscontent}}{}


\dropcap{I}n crystalline solids, the combination of spin-orbit coupling and crystal electric fields places strong
constraints on the shape of localized electronic wavefunctions \cite{fazekas_book}. The quantum mechanically defined multipole moments
provide a useful measure of the resulting complex angular distribution of the magnetization and charge densities \cite{multupole_rev_1, multupole_rev_2}.
Most conventional broken symmetry phases in solids involve the magnetic dipole moment of the electron. Remarkably, a 
large class of strongly correlated electron materials display nontrivial higher order multipolar moments, e.g., quadrupolar or octupolar moments, whose fluctuations and ordering
leads to a rich variety of phases such as quadrupolar 
heavy Fermi liquids \cite{sn_kondo_ti, review_exotic_multipolar, coleman_2015, floquet_hf}, superconductivity \cite{super_fq_pr_ti, pressure_hfm_super_quad_pr_ti, hfm_superconductivity_v, Sb_Nuclear_Quadrupole_1, Sb_Nuclear_Quadrupole_2}, and unusual multipolar symmetry-broken phases \cite{multupole_rev_1, multupole_rev_2, multupole_rev_3, shiina_multipole_1, shiina_multipole_2, kiss_thesis_2004, kiss_octupole_U,kiss_octupole_mag_2003}. 

While multipolar ordered phases fall under the purview of the celebrated Landau paradigm of symmetry-broken phases, they  have been termed as so-called `hidden orders':
mysterious phases of matter whose orderings are \textit{invisible} to conventional local probes (such as neutron scattering or magnetic resonance), but are remarkably still known to exist as their onset triggers non-analytic signatures in thermodynamic measurements \cite{pr_afq, cef_splitting,pr_nfl,mag_specific_heat_v, super_afq_rh}. 
\textcolor{black}{Studying the mysterious ordering patterns
of higher order multipoles is also often rendered challenging since they typically coexist with conventional dipolar moments}. Examples of such symmetry breaking which are of great interest include spin-nematic order \cite{Podolsky_demler_2005} in spin $S \geq 1$ quantum magnets, quadrupolar charge order in transition metal oxides,
and higher multipolar order in $f$-electron heavy fermion materials \cite{sblee_gyrotropy} such as URu$_2$Si$_2$ \cite{Chandra_2002, Chandra_2003, Tripathi_2005, Tripathi_2007, Santander_2009, Kotliar_Haule_2009, Kotliar_Haule_2010, Kotliar_Haule_2011, Okazaki_2011, rau_2012} and UBe$_{13}$ \cite{Stewart_1984, Cox_1987, Cox_1998}.
The quest to probe such orders has led to novel experimental techniques, e.g., elastoresistivity \cite{Fisher_2013, Fisher_2015, Fisher_2017} to elucidate the 
quadrupolar order associated with orbital nematicity in the iron pnictides.
A broad understanding of the nature of these symmetry broken phases, and
means to definitively demonstrate their existence, has proven to be a challenging, yet stimulating, endeavor for both theory and experiment.

\textcolor{black}{More recently, experiments have begun to study angle-dependent magnetostriction, the change in sample length induced by a magnetic field which can point along various
crystalline directions, in a wide class of materials with 
multipolar degrees of freedom \cite{sakai_mpipks_2018}.
Motivated by these experiments, in this work we theoretically discuss how magnetostriction provides 
a novel means to directly probe multipolar order parameters.}
The central observation of this paper is that an applied magnetic field allows for a {\it linear} coupling between lattice strain fields and a uniform octupole moment
which depends strongly on the applied field direction. In the absence of a dipolar moment, this enables measurements of the
magnetostriction to directly reveal the hidden octupolar order parameter.
We investigate such field-scaling behaviour of the magnetostriction for various magnetic field directions
by employing a symmetry-based Landau theory, which allows us to highlight the universal aspects of the physics and to show that this idea is broadly applicable
to a wide class of materials.

\textcolor{black}{Our work is motivated by a recent series of beautiful experiments on the Pr-based cage compounds belonging to the Pr(Ti,V,Ir)$_2$(Al,Zn)$_{20}$ family which 
form an ideal setting to study multipolar moments and associated hidden orders} \cite{pr_ti_v_super_mag, hfm_superconductivity_v, super_afq_rh, ir_super_afq, sbl_ybk_ap_2018}.
In these systems, the $4f^2$ electrons of Pr$^{3+}$ ions subject to CEFs host a ground non-Kramers doublet with solely higher-order moments: quadrupoles ($\mathcal{O}_{20}$ and $\mathcal{O}_{22}$) and octupole ($\mathcal{T}_{xyz}$) \cite{pr_fq, pr_afq}. Uncovering and understanding the pattern of multipolar ordering across this family of materials has remained an important open problem. 

The nature of the quadrupolar ordering in these cage compounds has been indirectly examined with a few techniques \cite{pr_v_nfl, sn_magneto_new_2} such as ultrasound experiments \cite{sn_elastic_new, ultrasound_ir, ultrasound_rh,ultrasound_v} (indicating softening of elastic modulus at quadrupolar ordering temperature, $T_{\mathcal{Q}}$), as well as NMR measurements 
(where the magnetic field-induced dipole moment is strongly dependent on the underlying quadrupolar phase \cite{nmr_ti}). More recently, magnetostriction and thermal expansion strain experiments \cite{magnetostiction_expt_ir} have also lent themselves as possible probes to study the transitions and the underlying quadrupolar phase.
By contrast, the octupolar ordered state has continued to remain an elusive phase of matter, with only indirect hints of its existence from NMR \cite{Santini_2000} and $\mu$SR \cite{kopmann_1998} measurements, but as yet no direct probe to reveal its existence \cite{walstedt_2018}.

In this study, motivated by showing how magnetostriction behaves in the presence of quadrupolar and octupolar orders, we focus on a Landau theory which permits both antiferro-quadrupolar ordering (AF$\mathcal{Q}$)
and ferro-octupolar ordering (F$\mathcal{O}$).
We study the scaling behaviour of the relative length change of the system with respect to an applied magnetic field strength ($h$) along different field directions. 
Denoting the quadrupolar and octupolar transition temperatures as $T_{\mathcal{Q}}$ and $T_{\mathcal{O}}$ respectively, we consider three regimes: (i) the paramagnetic phase above both transition temperatures ($T>T_{\mathcal{Q}}, T_{\mathcal{O}}$), (ii) intermediate temperatures ($T_{\mathcal{O}} < T < T_{\mathcal{Q}}$) where the system exhibits pure quadrupolar order, and (iii) below both ordering temperatures ($T<T_{\mathcal{Q}}, T_{\mathcal{O}}$) where the system features coexisting quadrupolar and octupolar orders.

Our studies predict a linear-in-$h$ scaling behaviour for length changes for a magnetic field applied along the [111] direction 
for $T< T_{\mathcal{O}}$. The coefficient of the linear-in-$h$ term, i.e. the ``magnetostriction coefficient'', 
is directly proportional to the ordered ferrooctupolar moment, thus providing a clear and distinct means to directly probe this order parameter.

A quick way to see this result is to note that the elastic energy of a cubic crystal is given by 
\begin{align}
F_{\text{lattice}} &= \frac{c_{11}}{2} \left(\epsilon_{xx}^2 + \epsilon_{yy}^2 + \epsilon_{zz}^2 \right) + \frac{c_{44}}{2} \left( \epsilon_{xy}^2 + \epsilon_{yz} ^2 + \epsilon_{xz} ^2 \right) \nonumber \\
& + c_{12} \left( \epsilon_{xx} \epsilon_{yy} + \epsilon_{yy} \epsilon_{zz} + \epsilon_{zz} \epsilon_{xx} \right) \ ,
\end{align}
where $\epsilon_{ij}$ and $c_{ij}$  refer to components of the strain tensor and elastic modulus tensor, respectively. Knowing $\epsilon_{ij}$ determines the fractional length change along the $\hat{\ell}$-axis via
$(\Delta L /L)_{\hat \ell} = \sum_{ij} \epsilon_{ij} \hat{\ell}_i \hat{\ell}_j$. As discussed below, an applied magnetic field enables a linear coupling between 
the strain field and the time-reversal-odd ferrooctupolar moment, $m$,
via a term in the free energy $\Delta F = - g_{\cal O} m (\epsilon_{yz} h_x + \epsilon_{xz} h_y + \epsilon_{xy} h_z)$, with a coupling constant $g_{\cal O}$. 
Minimizing $F_{\rm lattice} + \Delta F$
with respect to the strain, we find $\epsilon_{xy} \propto (g_{\cal O}/c_{44}) m h_z$, and cyclically for $\epsilon_{yz}, \epsilon_{xz}$, while diagonal
components of the strain tensor vanish. For a [111] field, where
$h_i = h/\sqrt{3}$, this leads to $(\Delta L /L)_{(1,1,1)} = (\epsilon_{xy} + \epsilon_{yz}+\epsilon_{xz})/3$ and so $(\Delta L /L)_{(1,1,1)}  \propto 
(g_{\cal O}/c_{44}) m h$. This direct relation between the linear-in-$h$ magnetostriction coefficient
and the ferrooctupolar order parameter for a magnetic field along the [111] direction is one of the central results of our paper.
Furthermore, we predict a characteristic hysteresis in the octupolar moment and the associated parallel length 
change as a function of magnetic field, arising from the symmetry-allowed cubic-in-$h$ coupling of the magnetic field to the octupolar moment.
Very recent (unpublished) experiments on PrV$_2$Al$_{20}$ indeed
appear to find a hysteretic linear-in-field magnetostriction, for a [111] magnetic field, below a transition at $T^* \approx 0.65$K.
\textcolor{black}{Our theoretical results for magnetostriction in the presence of octupolar order 
thus lend strong support to the idea that these seminal experiments \cite{sakai_mpipks_2018} herald the first and unambiguous discovery of octupolar order.}

Table \ref{tab:summary_scaling_all} provides a complete summary of the scaling behaviour of a variety of length change directions under different magnetic field directions,
including the effect of octupolar as well as quadrupolar order parameters.
Our predictions are expected to aid the investigation and identification of multipolar moments, as well as provide key signatures that indicate the presence of specific multipolar 
ordering. In particular, since we show that magnetostriction provides a direct probe of the octupolar order parameter, future experimental studies of this observable may shed light on the thermal and quantum critical behaviour associated with octupolar ordering in these compounds and a wide range of other materials.


\section{Landau Theory of Multipolar order} \label{sec_landau}

We present in this section, for the sake of self-containedness and to specify our notation, the Landau theory of multipolar order first introduced in Ref. \cite{sbl_ybk_landau_2018}.

The $4f^2$ electrons of Pr$^{3+}$ ions in the family of rare-earth metallic compounds Pr(Ti,V,Ir)$_2$(Al,Zn)$_{20}$ reside on a diamond lattice of cubic space group Fd$\bar{3}$m. Surrounding each Pr$^{3+}$ ion is a Frank-Kasper (FK) cage (16 Al atom polyhedra). The crystalline electric field (CEF) of this FK cage, with $T_d$ point group symmetry, splits the $J=4$ multiplet of the $4f^2$ electrons.
The ground states are experimentally found to form a non-Kramers doublet written in $\ket{J_z}$ basis as
\begin{equation}
\begin{aligned}
&\Gamma_{3} ^{(1)} = \frac{1}{2} \sqrt{\frac{7}{6}} \ket{4} - \frac{1}{2} \sqrt{\frac{5}{3}} \ket{0} + \frac{1}{2} \sqrt{ \frac{7}{6} } \ket{-4}, \\
&\Gamma_3 ^{(2)} = \frac{1}{\sqrt{2}} \ket{2} + \frac{1}{\sqrt{2}} \ket{-2}.
\end{aligned}
\end{equation}
These non-Kramers doublets transform as basis states of the $\Gamma_{3g}$ irrep. of $T_d$; here the subscript $g$($erade$) and $u$($ngerade$) denote even and odd under time-reversal, respectively. Moreover, this doublet is energetically well separated from the excited states, and so for energies much lower than this gap ($\gtrsim 50$K \cite{sn_kondo_ti}), the $\Gamma_{3g}$ doublets form an ideal basis to describe the low energy degrees of freedom. 

The $\Gamma_{3g}$ doublets can give rise to time-reversal even quadrupolar moments $\mathcal{O}_{22} = \frac{\sqrt{3}}{2} (J_x^2 - J_y^2)$ and $\mathcal{O}_{20} = \frac{1}{2} (2J_z^2 - J_x^2 - J_y^2)$ which
transform as $\Gamma_{3g}$, as well as a time-reversal odd octupolar moment ${\cal T}_{xyz} = \frac{\sqrt{15}}{6} \overline{J_x J_y J_z}$ which transforms as $\Gamma_{2u}$ (where the overline represents the fully symmetrized product). 
This can be seen from the group theory decomposition,
\begin{align}
\Gamma_{3g} \otimes \Gamma_{3g}  = \Gamma_{1g} \oplus \Gamma_{2u} \oplus \Gamma_{3g} 
\end{align}

Constructing a pseudospin basis ($ \{ \ket{\uparrow}, \ket{\downarrow} \} $) from the $\Gamma_{3g}$ doublets as
\begin{equation}
\begin{aligned}
&\ket{\uparrow} = \frac{1}{\sqrt{2}} [ \ket{\Gamma_{3} ^{(1)}} + i \ket{\Gamma_{3} ^{(2)}} ], \\
&\ket{\downarrow} = \frac{1}{\sqrt{2}} [ i \ket{\Gamma_{3} ^{(1)}} + \ket{\Gamma_{3} ^{(2)}} ]
\end{aligned}
\end{equation}
allows the multipolar moments to be neatly denoted by an effective pseudospin-1/2 operator $\vec{\tau} = (\tau^x, \tau^y, \tau^z)$
\begin{equation}
\begin{aligned}
&\tau^x = -\frac{1}{4}\mathcal{O}_{22}, ~~~ \tau^y = -\frac{1}{4}\mathcal{O}_{20}, ~~~ \tau^z = \frac{1}{3 \sqrt{5}}\mathcal{T}_{xyz}. 
\end{aligned}
\end{equation}
The perpendicular component of the pseudospin vector $\vec{\tau}^{\perp} \equiv (\tau^x, \tau^y)$ denotes the quadrupole moments, while $\tau^z$ denotes the octupolar moment. 
We also define the raising/lowering pseudospin operators $\tau^{\pm} = \tau^x \pm i \tau^y$. 

The ordering of these multipolar degrees of freedom acts as a mean field on the pseudospins, and breaks the degeneracy of the non-Kramers doublet.
In order to describe these pseudospin-symmetry broken phases, we resort to a Landau theory approach, focussing on the following 
order parameters,
\begin{equation}
\begin{aligned}
& \phi &&\equiv \langle \tau^+_{A} \rangle + \langle \tau^+_{B} \rangle , \\ 
& \tilde{\phi} &&\equiv \langle \tau^+_{A} \rangle - \langle \tau^+_{B} \rangle , \\ 
& m &&\equiv \langle \tau^z_{A} \rangle + \langle \tau^z_{B} \rangle , \\
& \tilde{m} &&\equiv \langle \tau^z_{A} \rangle - \langle \tau^z_{B} \rangle   ,
\label{eq:order_params_defined}
\end{aligned}
\end{equation}
Here, angular brackets $\langle ... \rangle$ denote thermal averages, while the $A,B$ subscripts denote the two sublattices of the diamond lattice.
The complex scalars $\phi$ and  $\tilde{\phi}$ describe ferroquadrupolar (F$\mathcal{Q}$) and anti-ferroquadrupolar (AF$\mathcal{Q}$) orders respectively, while the
real scalars $m$ and $\tilde{m}$ denote the ferrooctupolar (F$\mathcal{O}$) and anti-ferrooctupolar (AF$\mathcal{O}$) order parameters.

The local $T_{d}$ symmetry instilled by the FK cage provides a constraint on the possible terms permitted in the Landau theory. The generating elements of $T_{d}$ are $\mathcal{S}_{4z}$ (improper rotation of $\pi/2$ about the $\hat{z}$-axis) and $\mathcal{C}_{31}$ (rotation of $2 \pi /3$ about the body diagonal [111] axis). In addition to these point group symmetries, we also require that the terms in the Landau theory be invariant under spatial inversion about the diamond bond centre $\mathcal{I}$ (which swaps the $A$ and $B$ sublattices), as well as time-reversal $\Theta$. The behaviour of the multipolar moments under these symmetry constraints is detailed in Table \ref{tab_order} in SI \ref{app_multipolar_symm}.

In this work, we focus on a system where the primary order parameters are AF$\mathcal{Q}$ and F$\mathcal{O}$. As discussed in previous work \cite{sbl_ybk_landau_2018,hattori_afq_fq_2014},
the Landau theory of a system with  AF$\mathcal{Q}$ order necessarily admits a `parasitic'  secondary order parameter F$\mathcal{Q}$. Such mixing does not occur for the octupolar order parameter; motivated by 
explaining experiments on PrV$_2$Al$_{20}$ \cite{sakai_mpipks_2018}, we choose to work with only F$\mathcal{O}$ order and ignore the  AF$\mathcal{O}$ order parameter. We thus construct our
Landau theory using the order parameters $\phi$, $\tilde{\phi}$, and $m$.

\subsection{Interacting multipolar orders}

Equipped with the symmetry knowledge from Table \ref{tab_order} we can now write down the Landau free energy for this particular multipolar ordered system as
\begin{equation}
\begin{aligned}
F_{\mathcal{Q}, \mathcal{O}}[\phi, \tilde{\phi}, m] = F_{\tilde{\phi}} + F_{m} + F_{\phi} + F_{\tilde{\phi}, \phi, m} \ .
\label{landau_free_multi}
\end{aligned}
\end{equation}
Here, the free energies $F_{\tilde{\phi}}$, $F_{m}$, and $F_{\phi}$ denote the independent free energies of the AF$\mathcal{Q}$, F$\mathcal{O}$, and F$\mathcal{Q}$ orders.
Setting $\tilde{\phi} = | \phi | e^{i \tilde{\alpha}}$ and $\phi = | \phi | e^{i \alpha}$, we get
\begin{align}
 F_{\tilde{\phi}} &= \left[ \frac{{t}_{\tilde{\phi}}}{2} | {\tilde{\phi}}|^2 + {u}_{\tilde{\phi}} |\tilde{\phi}|^4 \right] + \left(  {l}_{\tilde{\phi}} + {w}_{\tilde{\phi}} \cos(6 \tilde{\alpha}) \right) |\tilde{\phi}|^6, \label{eq_afq_own} \\
 F_{m} &= \left[\frac{t_{m}}{2} m ^2 + u_m m^4 \right],  \label{eq_fo_own} \\
 F_{\phi} &= \left[\frac{{t}_{\phi}}{2} |{\phi}|^2 + {u}_\phi |{\phi}|^4 \right] + v_{\phi} \sin(3 \alpha) |\phi|^3, \label{eq_fq_own} 
\end{align}
The first two terms in \crefrange{eq_afq_own}{eq_fq_own},
in square brackets, are the usual mass and quartic interaction terms for AF$\mathcal{Q}$, F$\mathcal{O}$ and F$\mathcal{Q}$ order parameters.
We will choose $t_{\tilde{\phi}} = \textcolor{black}{(T - T_{\cal Q})/T_{\cal Q}}$, and $t_{m} = \textcolor{black}{(T - T_{\cal O}^{(0)})/T_{\cal O}^{(0)}}$ with $T_{\cal O}^{(0)} < T_{\cal Q}$, where $T$ denotes the
temperature. Focussing on the mass term alone, 
decreasing $T$ will thus lead to an anti-ferroquadrupolar order for $T < T_{\cal Q}$, and a lower temperature transition into a state with
coexisting ferro-octupolar order when $T < T_{\cal O}^{(0)}$.  These (bare) 
transition temperatures will be affected by the interplay of the two order parameters; in particular, the true octupolar
transition $T_{\cal O}$ will be renormalized from its bare value $T_{\cal O}^{(0)}$ due to the onset of quadrupolar order (besides fluctuation effects which we do not
consider here).
A measure of how close the two transition temperatures 
are to each other is provided by the ratio $(T_{\cal Q}-T_{\cal O})/(T_{\cal Q}+T_{\cal O})$. Finally, since  F$\mathcal{Q}$ is not considered to be a primary order parameter, we choose a \textcolor{black}{large positive mass term, $t_{\phi}$.}
The remaining non-trivial terms in Eqns.~\ref{eq_afq_own} and \ref{eq_fq_own} are the unusual sixth order and cubic ``clock'' terms, with respective
coefficients ${w}_{\tilde{\phi}}$  and $ v_{\phi} $, which fix the phases
of the AF${\mathcal{Q}}$ and F${\mathcal{Q}}$ order parameters. We set $ l_{\tilde{\phi}} > | w_{\tilde{\phi}} |$ to ensure that the free energy is bounded from below.

\begin{figure} [t]
\centering
  \includegraphics[width=0.85\linewidth]{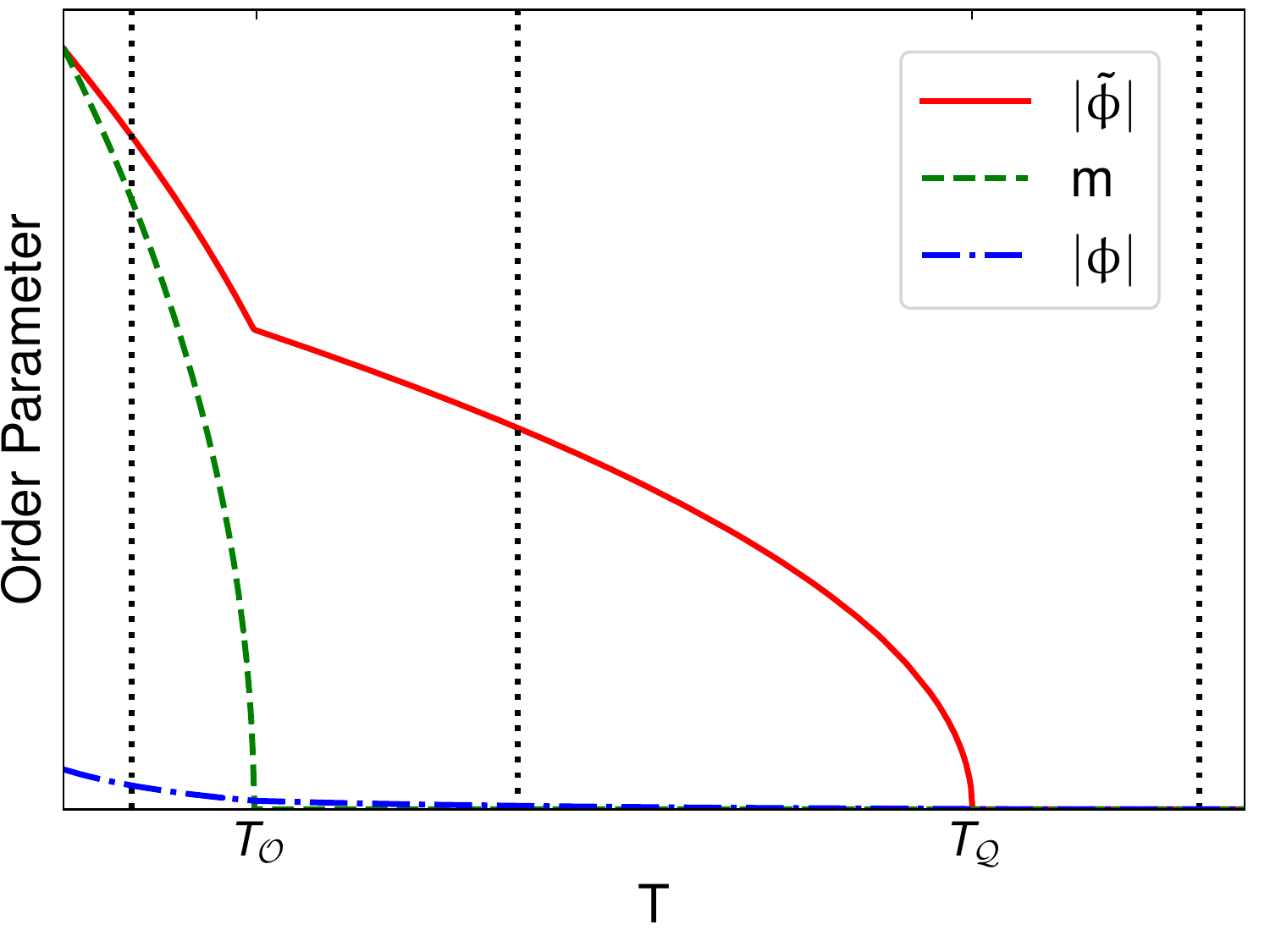}
  \caption{Phase diagram at zero magnetic field [$h=0$]. The temperature regimes studied in Sec. \ref{relative_length_all} are denoted by dashed lines at: $T<T_\mathcal{Q}, T_\mathcal{O}$, $T_\mathcal{O}<T<T_\mathcal{Q}$,  and $T>T_\mathcal{Q}, T_\mathcal{O}$. The order parameters for AF$\mathcal{Q}$, F$\mathcal{O}$, and F$\mathcal{Q}$ are denoted by $| \tilde{\phi} |$, $m$, and $| \phi |$, respectively. Here the  critical temperatures are $T_\mathcal{Q}$ and $T_\mathcal{O}$ [shifted from the bare $ T_{\mathcal{O}}^{(0)}$ due to the couplings $u_{\tilde{\phi}m}$ and $u_{\phi m}$ discussed in the main text].}
  \label{phase_diagram_0}
\end{figure}

The couplings between the different multipolar order parameters are encapsulated in $F_{\tilde{\phi}, \phi, m}$, namely between AF$\mathcal{Q}$ and F$\mathcal{Q}$ moments $(g_1, g_2)$, and between the quadrupolar and the octupolar moments $(u_{\phi m}, u_{\tilde{\phi}, m})$
\begin{equation}
\begin{aligned}
F_{\tilde{\phi}, \phi, m} &= g_1 |\phi | |\tilde{\phi} |^2 \sin(\alpha + 2\tilde{\alpha}) + g_2 | \tilde{\phi} |^2 | {\phi} |^2 \\
& +  {u}_{\phi m}  |{\phi}|^2 {m}^2 + {u}_{\tilde{\phi} {{m}}}  |\tilde{\phi}|^2 {m}^2,
\label{eq_afq_fo_fq_all}
\end{aligned}
\end{equation}
where the term $g_1$ is a symmetry-allowed cubic term.
We present in Fig. \ref{phase_diagram_0} the zero magnetic field phase diagram depicting both quadrupolar and octupolar transitions; with two primary order parameters AF$\mathcal{Q}$ (and its accompanying parasitic F$\mathcal{Q}$ moment) and F$\mathcal{O}$ ordering at critical temperatures of $T_{\mathcal{Q}}$ and $T_{\mathcal{O}}$, respectively. The octupolar transition temperature is shifted to $T_{\cal O}$, from its bare critical temperature $T_{\mathcal{O}}^{(0)}$,
due to the coupling of F$\mathcal{O}$ to AF$\mathcal{Q}$ and F$\mathcal{Q}$ via $u_{\tilde{\phi}m}$ and $u_{\phi m}$, respectively. The `kink' in the AF$\mathcal{Q}$ (as well as F$\mathcal{Q}$) at the octupolar ordering temperature reflects the non-analytic behaviour of the octupolar moment at its critical temperature. The dotted vertical lines denote specific temperature regions studied in Sec. \ref{relative_length_all}.

\subsection{Coupling of magnetic field to multipolar moments}

In order to study magnetostriction, it is important to understand how the magnetic field couples to the multipole moments.
Due to the lack of magnetic dipole moment supported by the $\Gamma_{3g}$ doublet, the magnetic field does not couple linearly to the states. One can derive the low energy 
magnetic field Hamiltonian by performing second-order perturbation theory in $\vec{h} \cdot \vec{J}$, where the low energy subspace is spanned by the $\Gamma_{3g}$ doublet, 
and the high energy subspace is spanned by the excited triplets $\Gamma_{4,5}$. This leads to
\begin{align}
\mathcal{H}_{\text{eff}} &= \gamma_0 \left[ \frac{\sqrt{3}}{2} (h_x^2 - h_y^2) \tau^x + \frac{1}{2}(3h_z^2 - h^2) \tau^y \right] \nonumber \\ 
			  & = \psi_H ^* \tau^+ +  \psi_H \tau^- \sim \psi_H^* \phi + \psi_H \phi^* \ .
\label{mag_micro}
\end{align}
In the above Eq. \ref{mag_micro}, $\vec{h} = (h_x, h_y, h_z)$ with $|\vec h| = h$, and $\gamma_0 \equiv \frac{-14}{3 \Delta(\Gamma_4)} + \frac{2}{\Delta(\Gamma_5)}$, where $\Delta(\Gamma_{4}), \Delta(\Gamma_{5})$ are the gaps between the low energy doublets and the corresponding triplet states at zero magnetic field. The effective coupling to
the ferroquadrupolar order is via $\psi_H \equiv \frac{\gamma_0 \sqrt{3}}{4} (h_x^2 - h_y^2) + i \frac{\gamma_0}{4}(3h_z^2 - h^2)$.
Based on the form of the coupling in Eq. \ref{mag_micro}, we infer that $\psi_H$ transforms identically to $\phi$ under the relevant symmetries.
Going to third-order in perturbation theory leads to a further ${\cal O}(h^3)$ coupling of the magnetic field to octupole moment
of the form $\sim h_x h_y h_z \tau^z$. 

Thus, the symmetry allowed effective magnetic field coupling to the quadrupolar moments is
\begin{equation}
\begin{aligned}
F_{\text{mag}}[\phi, \tilde{\phi}]  & =  \tilde{r}_H \sin(\theta_H + 2 \tilde{\alpha}) |\tilde{\phi}|^2  |\psi_H|  \\ 
& +  {r}_H \cos(\alpha - \theta_H) |{\phi}|   |\psi_H|  \\ 
& + \left( \tilde{s}_H  | \tilde{ \phi} |^2 + {s}_H  | { \phi} |^2 \right ) h^2 ,
\label{f_mag}
\end{aligned}
\end{equation}
where $|\psi_H| = \frac{\gamma_0}{4} \sqrt{  3(h_x^2 - h_y^2)^2 + (3h_z^2 - h^2)^2   }$, and $\tan(\theta_H) =  \frac{1}{\sqrt{3}}\frac{3 h_z^2 - h^2}{(h_x^2 - h_y^2) }$. The first (second) line in Eq. \ref{f_mag} is the symmetry allowed coupling to the AF$\mathcal{Q}$ (F$\mathcal{Q}$). The third line involves couplings permitted due to pure symmetry reasons that renormalize the mass terms of the AF$\mathcal{Q}$ and F$\mathcal{Q}$. Physically they arise from conduction electron mediated magnetic couplings (having integrated out the conduction electrons); similar coupling to the octupolar moment is also permitted [$\sim h^2 m^2$], which is formally introduced in Sec. \ref{sec_strain} via the magnetic field assisted coupling of the octupolar moment to the lattice strain. In the subsequent sections, we discuss magnetic fields applied along the [100] , [110] and [111] directions. For clarity, we present the value for $|\psi_H|$ and $\theta_H$ for the magnetic field directions discussed in subsequent sections, in Table \ref{tab:mag_directions}.
\begin{table}[h]
\begin{tabular}{ c | >{\centering\arraybackslash}m{0.9in} | >{\centering\arraybackslash}m{0.9in}}
\toprule
Magnetic Field, $\vec{h} = h \ \hat{n}$ & $|\psi_H| $& $\theta_H$   \\
\hline 
$\hat{n} = [100]$ & $  \frac{\gamma_0}{2} h^2$ & $-\pi/6$ \\
$\hat{n} = \frac{1}{\sqrt{2}} [110]$ & $ \frac{\gamma_0}{4} h^2$ & $-\pi/2$ \\
$\hat{n} = \frac{1}{\sqrt{3}} [111]$ & $0$ & $-$ \\
\bottomrule
\end{tabular}
\caption{Effective magnetic field strengths $|\psi_H|$, and associated complex angle $\theta_H$. For the $\hat{n} = \frac{1}{\sqrt{3}} [111]$, the magnetic field does not \textit{directly} couple to the quadrupolar moments, but can do so via $\tilde{s}_H$ and ${s}_H$, as described in the main text.}
\label{tab:mag_directions}
\end{table}

\section{Cubic Crystal Normal modes, and Relative Length Change Expression} \label{sec_cubic}

In this section, we introduce the expression of the free energy of a deformed face-centred cubic lattice, as well as its associated normal modes. We also formulate the relative length change expression in terms of the elastic strain components.

\subsection{Elastic energy of a cubic crystal}
In the spirit of Landau and Lifshitz, the general form of the free energy of a cubic crystal is constrained by the octahedral symmetry, $\mathcal{O}_h$, to be \cite{Landau_lifshitz_elasticity, luthi_2006} 
\begin{align}
F_{\text{lattice}} &= \frac{c_{11}}{2} \left(\epsilon_{xx}^2 + \epsilon_{yy}^2 + \epsilon_{zz}^2 \right) + \frac{c_{44}}{2} \left( \epsilon_{xy}^2 + \epsilon_{yz} ^2 + \epsilon_{xz} ^2 \right) \nonumber \\
& + c_{12} \left( \epsilon_{xx} \epsilon_{yy} + \epsilon_{yy} \epsilon_{zz} + \epsilon_{zz} \epsilon_{xx} \right) \ ,
\end{align}
where the crystal's deformation is described by the components of the strain tensor $\epsilon_{ik}$, and $c_{ij}$ is the elastic modulus tensor describing the stiffness of the crystal. Note that we use the common abbreviation of the elastic modulus tensor's indices i.e. $c_{xxxx} \equiv c_{11}$, $c_{xx yy} \equiv c_{12}$, $c_{xy xy} \equiv c_{44}$. This expression can be more elegantly written in terms of the normal modes of the cubic lattice, namely,
\begin{equation}
\begin{aligned}
F_{\text{lattice}} &= \frac{c_{B}}{2} \left(\epsilon_B^2 \right) + \frac{c_{11} - c_{12}}{2} \left( \epsilon_{\mu} ^2 + \epsilon_{\nu} ^2 \right)  \\
&+ \frac{c_{44}}{2} \left( \epsilon_{xy}^2 + \epsilon_{yz} ^2 + \epsilon_{xz} ^2 \right) \ ,
\label{F_lattice_normal}
\end{aligned}
\end{equation}
where $c_B$ is the bulk modulus,  $\epsilon_B \equiv \epsilon_{xx} + \epsilon_{yy} + \epsilon_{zz}$ is the volume expansion of the crystal, $\epsilon_{\nu} \equiv (2 \epsilon_{zz} - \epsilon_{xx} - \epsilon_{yy}) / \sqrt{3}$ and $\epsilon_{\mu} \equiv (\epsilon_{xx} - \epsilon_{yy})$ are lattice strains that transform as the $\Gamma_{3g}$ irrep. of the $O_h$ group, and the off-diagonal strain components transform as the $\Gamma_{5g}$ irrep. of $O_h$ group; here the subscript $g$ indicates even under time-reversal and spatial inversion (parity). We henceforth use Eq. \ref{F_lattice_normal} for the cubic crystal's elastic energy.

\subsection{General expression for relative length change}
The relative length change, $\Delta L / L$, of the crystal can be shown to be related to the components of the strain tensor, as described in more detail in SI \ref{app_length_derivation}. The general expression of the length change along a direction $\vec{\ell}$ is 
\begin{equation}
\left(\frac{\Delta L }{L}\right)_{\vec{\ell}} = \sum_{i,j=1} ^{3} \epsilon_{ij} \hat{\ell}_i \hat{\ell}_j  \ ,
\label{eq:len}
\end{equation}
where $\epsilon_{ij} \equiv \frac{1}{2}\left (\frac{\partial u_i}{\partial x_j} + \frac{\partial u_j}{\partial x_i} \right) $ is the familiar strain tensor, and $\hat{\ell}_i$ is the $i^{th}$ component of unit vector $\hat{\ell}$. For ease of understanding the strain tensor, we use the convention that $\epsilon_{11} = \epsilon_{xx}$, $\epsilon_{12} = \epsilon_{xy}$ etc. We apply Eq. \ref{eq:len} to particular length change directions in Sec. \ref{relative_length_all}.

\section{Symmetry Allowed Coupling of Multipolar Moments and Cubic Crystal Normal Modes} \label{sec_strain}

We now turn our attention to the problem of coupling the lattice normal modes of the cubic crystal to the multipolar moments. We recall that the cubic crystal structure supports macroscopic normal modes that transform as irreps. of $O_h$, while the Landau free energy of the multipolar moments ($F$) is constructed subject to symmetries of the local $T_d$ environment. 
The symmetry constraints on $F$ ensure that in principle \textit{only select} normal modes of the crystal that transform as the irreps. of $T_d$ are permitted to couple to the multipolar moments. In the present case, \textit{all} the cubic normal modes presented in Eq. \ref{F_lattice_normal} also transform as irreps. under $T_d$ (as can be explicitly verified), and so \textit{all} of the aforementioned strain modes can participate in the coupling.
In the next two subsections, we consider the direct coupling of quadrupolar moments to the cubic normal modes, and then tackle the magnetic field assisted octupolar coupling to the lattice normal modes.

\subsection{Coupling of quadrupolar moment to lattice strain}

Coupling between the quadrupolar moments and the lattice normal modes appears as a natural choice, as the quadrupolar moments and the lattice strains are both even under time-reversal. Moreover, both the normal modes $\{ \epsilon_{\mu}, \epsilon_{\nu} \}$ \textit{and} the quadrupolar moments $\{ \mathcal{O}_{22}, \mathcal{O}_{20} \}$ transform as $\Gamma_{3g}$ irreps. of $T_d$ (the aforementioned lattice normal modes also transform as $\Gamma_{3g}$ in $O_h$, as $T_d$ is a subgroup of $O_{h}$). This similarity in how they transform under $T_d$ allows a linear coupling between the aforesaid lattice normal modes and quadrupolar moments. Thus, the Landau free energy of the multipolar moments shown in Eqs. \ref{landau_free_multi}, \ref{f_mag} gets augmented by the following lattice elastic energy and coupling terms to quadrupolar moments,
\begin{align}
F_{\text{strain}, \mathcal{Q}} [\tilde{\phi}, \phi, {\epsilon_{\mu, \nu}}]  & = \frac{c_{11} - c_{12}}{2} \left( \epsilon_{\mu} ^2 + \epsilon_{\nu} ^2 \right) \label{f_strain_quad}   \\ 
& -g_{\mathcal{Q}} \epsilon_{\mu} \left[  \langle \tau^x_A\rangle +  \langle \tau^x_B\rangle  \right]  -g_{\mathcal{Q}} \epsilon_{\nu} \left[   \langle \tau^y_A\rangle +  \langle \tau^y_B\rangle     \right] \nonumber \ , 
\end{align}
where $g_{\mathcal{Q}}$ is the coefficient of coupling between the quadrupolar moments and lattice strain tensors. Note that we include the coupling of the lattice strain to the quadrupole moment on each sublattice. Using the definition of the order parameter $\phi$ defined in Eq. \ref{eq:order_params_defined}, and minimizing with respect to the lattice degrees of freedom ($\frac{\delta F_{strain, \mathcal{Q}}}{\delta \epsilon_{\mu}}$ etc.)  yields the total strain for each normal mode
\begin{equation}
\begin{aligned}
& \epsilon_{\mu} =   \frac{g_{\mathcal{Q}} }{(c_{11} - c_{12})}|\phi| \cos{\alpha}  \ ,  \\
& \epsilon_{\nu}  =   \frac{g_{\mathcal{Q}} }{(c_{11} - c_{12})}|\phi| \sin{\alpha}  \ .
\label{eq:epsilon_mu_nu}
\end{aligned}
\end{equation} 
Substituting the expression for the minimized lattice strains from Eqs. \ref{eq:epsilon_mu_nu} back into Eq. \ref{f_strain_quad}, yields
\begin{equation}
F_{\text{strain}, \mathcal{Q}}  [ \tilde{\phi}, \phi ] = - \frac{g_{\mathcal{Q}}^2}{2(c_{11} - c_{12})}  |\phi|^2  \ . \\
\label{f_quad_end}
\end{equation}
Thus, the coupling of the lattice degrees of freedom to the quadrupolar moments results in renormalizing the mass term of $\phi$.

\subsection{Coupling of octupolar moment to lattice strain} \label{sec_octupolar_strain_coupling}

A direct linear coupling between the octupolar moment $\mathcal{T}_{xyz}$ and the lattice normal modes is not permitted, as the octupolar moment is odd under time-reversal. However, this potential difficulty can be alleviated by the introduction of the time-reversal odd magnetic field $\vec{h}$ which assists in the coupling between the lattice degrees of freedom and octupolar moment. Thus, the Landau free energy of the multipolar moments shown in Eqs. \ref{landau_free_multi}, \ref{f_mag} gets augmented by the following lattice elastic energy and the coupling terms to the octupolar moments,
\begin{align}
F_{\text{strain}, \mathcal{O}}[m, \{\epsilon_{xy, yz, xz} & \}]  = \frac{c_{44}}{2} \left( \epsilon_{xy}^2 + \epsilon_{yz} ^2 + \epsilon_{xz} ^2 \right) \nonumber \\
& -g_{\mathcal{O}}  m \left[ h_x \epsilon_{yz} + h_y \epsilon_{xz} + h_z \epsilon_{xy} \right] \label{f_strain_oct} \\
& - \gamma \left[ h_x h_y \epsilon_{xy} + h_x h_z \epsilon_{xz} + h_y h_z \epsilon_{yz} \right] \nonumber , 
\end{align}
where we use the definition of $m$ from Eq. \ref{eq:order_params_defined}, and $g_{\mathcal{O}}$ is the coefficient of coupling between the octupolar moment and lattice strain. We also include another symmetry allowed direct coupling between the magnetic field and the same lattice normal modes (with proportionality constant $\gamma$, equivalent on both sublattices). Physically, this kind of term could arise from the independent coupling of the magnetic field and lattice strain to the conduction electrons (and after integrating out the conduction electrons); we discuss this matter briefly in Sec. \ref{conclusions_sec}.

Minimizing with respect to the lattice degrees of freedom yields the following expressions for the (total) lattice strains
\begin{equation}
\begin{aligned}
& \epsilon_{xy} = \left(\frac{g_{\mathcal{O}} h_z}{c_{44}} \right) m + \gamma \frac{h_x h_y}{c_{44}} \ ,  \\
& \epsilon_{xz} = \left(\frac{g_{\mathcal{O}} h_y}{c_{44}}\right) {m} + \gamma \frac{h_x h_z}{c_{44}} \ ,  \\
& \epsilon_{yz} = \left(\frac{g_{\mathcal{O}} h_x}{c_{44}}\right) {m} + \gamma \frac{h_y h_z}{c_{44}} \ . 
\label{eq:oct_length}
\end{aligned}
\end{equation}
Substituting the expression for the minimized lattice strains from Eqs. \ref{eq:oct_length} into Eq. \ref{f_strain_oct}, yields
\begin{equation}
\begin{aligned}
F_{\text{strain}, \mathcal{O}} [m] &= - \frac{g_{\mathcal{O}}^2}{2c_{44}} \left(h_x^2 + h_y ^2 + h_z^2 \right) m ^2 \\
& - \left(\frac{3 g_{\mathcal{O}} \gamma}{c_{44}} h_x h_y h_z\right) m + \mathcal{O}(h^4) \ . 
\label{f_oct_end}
\end{aligned}
\end{equation}
Thus, the coupling of the lattice degrees of freedom to the octupolar moment results in renormalizing the mass term of the octupolar moment quadratically in $h$; \textcolor{black}{it also introduces an $\mathcal{O}(h^3)$ coupling term between the octupolar moment and the magnetic field, which renormalizes the coefficient of the already present $h_x h_y h_z m$ from third-order in perturbation theory in $\vec{h} \cdot \vec{J}$.}

\section{Relative Length Change Under Magnetic Field Along Different Directions}\label{relative_length_all}

In this section, we examine the relative length change, $\Delta L / L$, for magnetic fields applied along $[100]$, $[110]$, $[111]$ directions and examine the scaling in magnetic field strength, $h$. For the sake of clarity, we write down the complete Landau theory, including the magnetic field couplings, and after having integrated out the lattice degrees of freedom (as discussed in the previous section, resulting in renormalizing the mass terms of the order parameters) 

\begin{equation}
\begin{aligned}
F[\phi, \tilde{\phi}, m]  & = F_{\mathcal{Q}, \mathcal{O}}[\phi, \tilde{\phi}, m] + F_{\text{mag}}[\phi, \tilde{\phi}] \\
& + F_{\text{strain}, \mathcal{Q}}  [ \tilde{\phi}, \phi ] + F_{\text{strain}, \mathcal{O}} [m]  \ , 
\label{landau_free_all}
\end{aligned}
\end{equation}
where $F_{\mathcal{Q}, \mathcal{O}}[\phi, \tilde{\phi}, m]$ is defined in Eq. \ref{landau_free_multi}, $F_{\text{mag}}[\phi, \tilde{\phi}]$ is defined in Eq. \ref{f_mag}, and the terms from the strain couplings $F_{\text{strain}, \mathcal{Q}}  [ \tilde{\phi}, \phi ]$ and  $F_{\text{strain}, \mathcal{O}} [m]$ are defined in Eqs. \ref{f_quad_end} and \ref{f_oct_end}, respectively. We present in SI \ref{app_landau_params}, the values of the Landau parameters chosen for the study conducted in this and the subsequent sections. For each magnetic field direction, we examine three temperature regimes, namely: above all critical temperatures, between the quadrupolar and octupolar critical temperatures, and below both critical temperatures. 
\pagebreak

The scaling relations can be inferred by substituting the expressions for the (extremized) strain in Eqs. \ref{eq:epsilon_mu_nu} and \ref{eq:oct_length} into Eq. \ref{eq:len} to yield the following expressions in Eqs. \ref{eq_length_subbed_multipolar}.
\begin{widetext}
\begin{align}
& \left(\frac{\Delta L }{L}\right)_{(1,0,0)} &&= \frac{\epsilon_B}{3}  - \frac{\epsilon_{\nu}}{2 \sqrt{3}}  + \frac{\epsilon_{\mu}}{2} &&=  \frac{1}{3} \epsilon_B + {\left[ -g_{\mathcal{Q}}  \frac{ \sin(\alpha) - \sqrt{3}\cos(\alpha)}{2 \sqrt{3} (c_{11} - c_{12})}\right]} | \phi |  \nonumber \ ,   \\ 
& \left(\frac{\Delta L }{L}\right)_{(0,1, \pm 1)}  &&= \frac{\epsilon_B}{3}  + \frac{\epsilon_{\nu}}{4 \sqrt{3}}  - \frac{\epsilon_{\mu}}{4}  \pm  \epsilon_{yz}  &&=  \frac{1}{3}\epsilon_B   + {\left[ g_{\mathcal{Q}}  \frac{ \sin(\alpha) - \sqrt{3}\cos(\alpha)}{4 \sqrt{3} (c_{11} - c_{12})}\right]} | \phi | \pm  \frac{g_\mathcal{O} m}{c_{44}}  h_x \pm  \frac{\gamma h_y h_z}{c_{44}}  \   \nonumber \\
& \left(\frac{\Delta L }{L}\right)_{(1,\pm 1,0)}  &&= \frac{\epsilon_B}{3} - \frac{\epsilon_\nu}{2\sqrt{3}} \pm  \epsilon_{xy}  &&=  \frac{1}{3}\epsilon_B   +  {\left[ -g_{\mathcal{Q}} \frac{\sin(\alpha) }{2 \sqrt{3} (c_{11} - c_{12})}  \right]} |\phi | \pm  \frac{g_\mathcal{O} m}{c_{44}}  h_z \pm  \frac{\gamma h_x h_y}{c_{44}}  \label{eq_length_subbed_multipolar}
 \ ,   \\ 
& \left(\frac{\Delta L }{L}\right)_{(1,\pm1,1)} &&= \frac{\epsilon_B}{3}    +  \frac{2 \left( \pm \epsilon_{xy}  \pm  \epsilon_{yz}  + \epsilon_{xz} \right)}{3} &&= \frac{1}{3}  \epsilon_B  +  \frac{2g_{\mathcal{O}} m }{3 c_{44}} \left[ \pm h_z \pm h_x + h_y  \right]  + \frac{2 \gamma}{3 c_{44}} \left[ \pm h_x h_y \pm h_y h_z + h_x h_z \right]  \nonumber , \\
& \left(\frac{\Delta L }{L}\right)_{(\mp1,1,\pm2)} &&= \frac{\epsilon_B}{3} + \frac{\epsilon_\nu}{2\sqrt{3}} + \frac{ \left( \mp \epsilon_{xy} \pm 2 \epsilon_{yz} - 2\epsilon_{xz}  \right)}{3}  && =  \frac{1}{3} \epsilon_B + {\left[ g_{\mathcal{Q}}  \frac{ \sin(\alpha) }{2\sqrt{3} (c_{11} - c_{12})}\right]} | \phi | +   \frac{g_\mathcal{O} m}{3c_{44}}  \left[ \mp h_z \pm 2h_x - 2h_y  \right] \nonumber \\
& & && & + \frac{\gamma}{3c_{44}} \left[ \mp h_x h_y \pm 2h_y h_z - 2 h_x h_z  \right] \nonumber \ .  
\end{align}
\end{widetext}
\noindent 
Here we have used the definition of the normal modes introduced in Eq. \ref{F_lattice_normal}, and define $\overline{{g}_{\mathcal{Q}} } \equiv \frac{g_{\mathcal{Q}} }{ \sqrt{3} (c_{11} - c_{12})} (...)$, wherein $(...)$ includes the complex-angle dependent terms in Eq. \ref{eq_length_subbed_multipolar}; this definition will be helpful in Table \ref{tab:summary_scaling_all} below. The exact form of which complex angle term is included ($\frac{ \sin(\alpha) }{2\sqrt{3} (c_{11} - c_{12})}$ or $\frac{ \sin(\alpha) - \sqrt{3}\cos(\alpha)}{2 \sqrt{3} (c_{11} - c_{12})}$) can be inferred from context i.e. the direction of length change examined $\vec{\ell}$ under particular magnetic field direction $\hat{n}$. The parasitic F$\mathcal{Q}$ moment is written as $| \phi | = \left( \phi_0 + \phi_{h} h^2 \right)$ due to the even-in-$h$ behaviour of the quadrupolar moment [as described later in the main text]. 
The constant ($\phi_0$) and quadratic ($\phi_{h}$) scaling-coefficients depend on the value of the Landau parameters as well as the temperature being probed; the quantitative value is thus not of great importance for the scaling behaviour. The key point to retain is that the value of these coefficients is small as compared to the conduction electron generated terms ($\sim \gamma$), reflecting the weak, parasitic nature of the F$\mathcal{Q}$ moment. 
We present in Table \ref{tab:summary_scaling_all} the scaling behaviours of length change parallel and perpendicular to the three primary magnetic field directions. 

\setlength{\extrarowheight}{1.2pt}
\begin{table*}[t]
\caption{Scaling relation for relative length change of system ${\Delta L}/{L}_{\vec{\ell}}$ along direction $\vec{\ell}$ for magnetic field applied along $\hat{n}$ direction. For each $\hat{n}$, we present the length change parallel and (the two) perpendicular directions with respect to $\hat{n}$. \textcolor{black}{F$\mathcal{Q}$ moment is expressed as $| \phi | = \left(  \phi_0  +  \phi_{h}  h^2 \right)$ due to the even-in-$h$ behaviour of the quadrupolar moment [described in main text]. $\phi_{0, {h}}$ are constants that arise from the parasitic F$\mathcal{Q}$ moment and are thus diminutive, as compared to the conduction electrons' term ($\sim \gamma / c_{44}$). The complex-angle (${\alpha}$) dependent parts of Eq. \ref{eq_length_subbed_multipolar} are included in the definition of the quadrupolar--lattice strain coupling, $\overline{g_{\mathcal{Q}}}$; the exact form of the complex angle term can be inferred from consulting Eq. \ref{eq_length_subbed_multipolar} for the appropriate $\hat{n}$ and $\vec{\ell}$ directions. The octupolar--lattice strain coupling is denoted by $g_{\mathcal{O}}$.}}
\begin{tabular*}{\hsize}{@{\extracolsep{\fill}}c|| c | c | c | c}
\toprule
\hline
Magnetic field  & \multicolumn{4}{c}{${\Delta L}/{L}_{\vec{\ell}}$ scaling}  \\
$\vec{h} = h \ \hat{n}$  & $\vec{\ell}$ & $T>T_{\mathcal{Q}}, T_{\mathcal{O}}$ & $T_{\mathcal{O} }< T < T_{\mathcal{Q}}$ & $T<T_{\mathcal{Q}}, T_{\mathcal{O}}$ \\
\hline
\multirow{3}{*}{$\hat{n} = [100]$} & $\vec{\ell} = (1,0,0)$ &  \multirow{1}{*}{ $\left(\overline{{g}_{\mathcal{Q}} }   \phi_{h}    \right)  h^2$} & \multirow{1}{*}{$  \overline{{g}_{\mathcal{Q}} }   \phi_0   + \left(\overline{{g}_{\mathcal{Q}} }    \phi_{h}   \right)  h^2 $} & \multirow{1}{*}{$  \overline{{g}_{\mathcal{Q}} }   \phi_0   + \left(\overline{{g}_{\mathcal{Q}} }    \phi_{h}   \right)  h^2 $}
\\
& \multirow{2}{*}{$\vec{\ell} = (0,1,\pm1)$} &  \multirow{2}{*}{ $\left(\overline{{g}_{\mathcal{Q}} }   \phi_{h}    \right)  h^2$} & \multirow{2}{*}{$  \overline{{g}_{\mathcal{Q}} }   \phi_0   + \left(\overline{{g}_{\mathcal{Q}} }    \phi_{h}   \right)  h^2 $} & \multirow{2}{*}{$  \overline{{g}_{\mathcal{Q}} }   \phi_0   \pm \left(\frac{g_{\mathcal{O}}}{c_{44}} m   \right) h + \left(\overline{{g}_{\mathcal{Q}} }    \phi_{h}   \right)  h^2 $}
\\
& & & & 
\\
\hline
\multirow{6}{*}{$\hat{n} = \frac{1}{\sqrt{2}} [110]$} & $\vec{\ell} = (1,1,0)$ & $  \left(\frac{\gamma}{2c_{44}} + \frac{\overline{{g}_{\mathcal{Q}} }}{2}   \phi_{h}   \right) h^2 $ & $\overline{{g}_{\mathcal{Q}} }   \phi_0    + \left(\frac{\gamma}{2c_{44}} + \frac{\overline{{g}_{\mathcal{Q}} }}{2}   \phi_{h}   \right) h^2 $ & $\overline{{g}_{\mathcal{Q}} }   \phi_0    + \left(\frac{\gamma}{2c_{44}} + \frac{\overline{{g}_{\mathcal{Q}} }}{2}   \phi_{h}   \right) h^2 $ 
\\ 
& & & &
\\
& $\vec{\ell} = (1,-1,1)$ & $ - \left(\frac{\gamma}{3c_{44}}\right) h^2 $ & $- \left(\frac{\gamma}{3c_{44}}\right) h^2$ & $- \left(\frac{\gamma}{3c_{44}}\right) h^2$
\\
& & & &
\\
& $\vec{\ell} = (-1,1,2)$ & $  \left(- \frac{\gamma}{6c_{44}} + \frac{\overline{{g}_{\mathcal{Q}} }}{2}   \phi_{h}   \right) h^2 $ & $\overline{{g}_{\mathcal{Q}} }   \phi_0   + \left(- \frac{\gamma}{6c_{44}} + \frac{\overline{{g}_{\mathcal{Q}} }}{2}   \phi_{h}   \right) h^2$ & $\overline{{g}_{\mathcal{Q}} }   \phi_0  + \left(- \frac{\gamma}{6c_{44}} + \frac{\overline{{g}_{\mathcal{Q}} }}{2}   \phi_{h}   \right) h^2$
\\
\hline
& & & &
\\
\multirow{4}{*}{$\hat{n} = \frac{1}{\sqrt{3}} [111]$} & $\vec{\ell} = (1,1,1)$ & $ \left(\frac{2\gamma}{3c_{44}}\right)  h^2$ & $\left( \frac{2\gamma}{3c_{44}} \right) h^2$ & $ \left(\frac{2 g_{\mathcal{O}}}{\sqrt{3} c_{44}}m\right) h + \left(\frac{2\gamma}{3c_{44}}\right) h^2$ \\
& & & &
\\
 & \multirow{1}{*}{$\vec{\ell} = (1,-1,0)$} &  \multirow{2}{*}{$\left( - \frac{\gamma}{3c_{44}} + \frac{\overline{{g}_{\mathcal{Q}} }}{3}   \phi_{h}   \right) h^2$}  & \multirow{2}{*}{$\overline{{g}_{\mathcal{Q}} }   \phi_0   + \left( -\frac{\gamma}{3c_{44}} + \frac{\overline{{g}_{\mathcal{Q}} }}{3}   \phi_{h}  \right) h^2$} & \multirow{2}{*}{$\overline{{g}_{\mathcal{Q}} }   \phi_0   - \left(\frac{g_{\mathcal{O}}}{\sqrt{3}c_{44}} m \right)h  + \left(- \frac{\gamma}{3c_{44}} + \frac{\overline{{g}_{\mathcal{Q}} }}{3}   \phi_{h}   \right) h^2$} \\
 & \multirow{1}{*}{$\vec{\ell} = (1,1,-2)$} & & & \\
 \hline
\bottomrule
\end{tabular*}
\label{tab:summary_scaling_all}
\end{table*}

The conclusions that can be drawn from Eq. \ref{eq_length_subbed_multipolar} and Table \ref{tab:summary_scaling_all} are striking. Firstly, in the zero field limit, we recall that there is a built-in three-fold degeneracy in the F$\mathcal{Q}$ due to the $v_{\phi}$ term in Eq. \ref{eq_fq_own}. One can consider the scenario of the zero field-limit being achieved by applying a magnetic field along [100] and tuning it to zero; this causes one of the three degenerate solutions to be chosen i.e. $\alpha = \frac{5 \pi}{6}$ at the zero-field limit [as seen in the next section]. With this particular $\alpha$-solution, the zero field $\left(\frac{\Delta L }{L}\right)_{(1,0,0)}$ is larger (i.e. more negative) than $\left(\frac{\Delta L }{L}\right)_{(1,\pm 1,0)}$. Secondly, the hitherto mysterious octupolar moment can now be determined (up to a proportionality constant) by measuring the slope of the linear-in-$h$ behaviour of the length change both parallel and perpendicular to magnetic fields applied along the $[111]$ direction; the linear behaviour is also apparent for perpendicular length changes to magnetic field applied along [100] direction. This provides a clear signature for the onset of the octupolar ordering as well as a means to study the general behaviour of the octupolar moment (up to a proportionality constant) with respect to other external variables such as temperature, $T$.
Thirdly, for magnetic fields applied along the $[111]$ direction (and for $T<T_{\mathcal{Q}}, T_{\mathcal{O}}$) the length change parallel to the magnetic field has (negative) twice the slope of the linear-in-$h$ term $\textit{and}$ (negative) twice the quadratic background as the length changes perpendicular to the field. This provides a distinct verification as to the validity of the theory.

In next subsections we elaborate on important details regarding the multipolar moments and examine their scaling-dependency on the magnetic field, to provide justification of the results presented in Table \ref{tab:summary_scaling_all}. Prior to discussing the particular magnetic field directions, we examine the consequences of our choice of parameters, which have the following properties 
\begin{equation}
\begin{aligned}
& v_\phi < 0, && w_{\tilde{\phi}} > 0, && g_1 < 0, && \tilde{r}_H  > 0, && {r}_H > 0 .
\end{aligned}
\end{equation}
In minimizing the free energy in Eq. \ref{landau_free_all}, we examine each of the complex angular-dependent terms and determine the value of the corresponding angles that minimize (magnetic field independent terms in) the Landau free energy
\begin{align}
& |{w}_{\tilde{\phi}}| | \tilde{\phi} |^{6} \cos(6 \tilde{\alpha})  && \overset{\text{min}[\tilde{\alpha}]}{\implies} \tilde{\alpha} = \frac{\pm \pi}{6}, \frac{\pm \pi}{2}, \frac{\pm 5 \pi}{6}   \nonumber  \\
& - |v_{\phi}| | \phi | ^3 \sin(3 \alpha) && \overset{\text{max}[{\alpha}]}{\implies} \alpha = \frac{-\pi}{2}, \frac{ \pi}{6}, \frac{5 \pi}{6} \\
& - |g_1| | \phi | |\tilde{\phi}|^2 \sin(\alpha + 2\tilde{\alpha})  && \overset{\text{max}[\alpha, \tilde{\alpha}]}{\implies} (\alpha, \tilde{\alpha}) = \left(\frac{5\pi}{6}, \frac{5\pi}{6}\right) \nonumber
\label{eq_min_main}
\end{align}
In the absence of external magnetic fields, there is a built-in degeneracy that allows a host of possible solutions, namely
\begin{align}
(\alpha, \tilde{\alpha}) =& (-\pi /2, \pm \pi/2), (\pi /6, -5 \pi/6), (\pi /6, \pi/6), \nonumber \\
&  (5\pi /6, - \pi/6), (5\pi /6, 5 \pi/6) \ .
\end{align}

\subsection{$\vec{H} \parallel [100]$} \label{100_angle}

For this magnetic field direction, we now examine the magnetic field dependent complex angular terms and determine the value of the corresponding angles that minimize the Landau free energy (having set $\theta_H = - \pi/6$),
\begin{align}
& | {r}_H | \cos(\alpha - \theta_H)  | \psi_H | |{\phi}| && \overset{\text{min}[{\alpha}]}{\implies} \tilde{\alpha} = \frac{-7\pi}{6}, \frac{5 \pi}{6}  \nonumber \\
&  |\tilde{r}_H| |\psi_H| |\tilde{\phi}|^2 \sin( - \frac{\pi}{6} + 2\alpha)  && \overset{\text{min}[\tilde{\alpha}]}{\implies} \alpha = \frac{-\pi}{6}, \frac{5 \pi}{6}  \label{eq_min_100}
\end{align}
Thus, the choice of $ (\alpha, \tilde{\alpha}) = (\frac{5\pi}{6}, \frac{5\pi}{6})$ extremizes all of the above expressions simultaneously. This result is the justification of the zero-field-limit conclusion stated above, where the $h=0$ limit is achieved by tuning down a magnetic field applied along the $[100]$ direction such that the $\alpha = 5 \pi/6$ solution is chosen. Now, the scaling behaviour of the quadrupole moment can be understood by deriving approximate analytical expressions; this is easiest to calculate for $T> T_{\mathcal{Q}}, T_{\mathcal{O}}$, and more tedious in the other two temperature regimes. The numerical solution of the order parameters in the three temperature regimes is presented in SI \ref{app_order_plot} Fig. \ref{100_magneto_all}.

For $T> T_{\mathcal{Q}}, T_{\mathcal{O}}$, $ | \tilde{\phi} |$ and $m$ are zero. Extremizing Eq. \ref{landau_free_all} with $\tilde{\phi} = m = 0$ [and knowledge of $\alpha = 5\pi/6$], and reasonably assuming that parasitic $\phi$ is small, we arrive at 
\begin{align}
| \phi | {\approx} \left(\frac{\gamma_0}{2}  |r_H|\right)  h^2 \ , 
\end{align}
where we have reintroduced $|\psi_H| = \frac{\gamma_0}{2} h^2$. In the notation of Table \ref{tab:summary_scaling_all}, here $ \phi_{h} = \frac{\gamma_0}{2}  |r_H| h^2$ for this magnetic field direction.

A similar (albeit more tedious) approach can be adapted to determine an approximate scaling behaviour in the other two temperature regimes of $T_{\mathcal{O}} < T < T_{\mathcal{Q}}$ (where $| \phi |, | \tilde{\phi} | \neq 0$ and $m=0$) and $T < T_{\mathcal{Q}}, T_{\mathcal{O}}$ (where all of $| \phi |, | \tilde{\phi}|, m  \neq 0$). Doing so, we arrive at the leading-order-in-$h$ scaling of $| \phi | \approx \big|_{T < T_{\mathcal{Q}}}  \phi_0  + \phi_{h} h^2$, where $  \phi_{0,h}  $ are a collection of constants. The constant shift $ \phi_{0} $ result is reasonable as below $T_{\mathcal{Q}}$ the AF$\mathcal{Q}$ is non-vanishing even at zero magnetic field, and thus the accompanying parasitic F$\mathcal{Q}$ is finite for zero magnetic field. We affirm these scaling behaviours by performing a thorough numerical study of the full Landau free energy in the temperature regimes of interest. As can be seen in SI \ref{app_order_plot} Fig. \ref{100_magneto_all}, the F$\mathcal{Q}$ moment is indeed an even function-in-$h$. Moreover, for $T< T_{\mathcal{Q}}$, the F$\mathcal{Q}$ moment is finite even at zero magnetic field due to the AF$\mathcal{Q}$ having spontaneously ordered. 

\subsection{$\vec{H} \parallel [110]$} \label{110_angle}

For magnetic fields along this direction, unlike the $[110]$ direction, there is a lack of harmony in the complex angles that minimize the Landau free energy. For the AF${\mathcal{Q}}$ and F${\mathcal{Q}}$ moments, the complex angles that extremize the respective magnetic field dependent terms are
\begin{equation}
\begin{aligned}
&  | \tilde{r}_H | \sin(\frac{-\pi}{2} + 2 \tilde{\alpha}) | \psi_H | |\tilde{\phi}|^2  && \overset{\text{min}[\tilde{\alpha}]}{\implies} \tilde{\alpha} =  0, \pm \pi \\
& | {r}_H | \cos(\alpha - (-\frac{\pi}{2}))  | \psi_H | |{\phi}| && \overset{\text{min}[{\alpha}]}{\implies} {\alpha} = \frac{\pi}{2} 
\end{aligned}
\end{equation}
This lack of harmony in the choice of polar angle leads to a competition between the two terms: the magnetic field dependent terms desire $\tilde{\alpha} = 0$ and ${\alpha} = \pi/2$; while we recall from Eq. \ref{eq_min_main} the anisotropic sextic (cubic) term desires $\tilde{\alpha} = \pm 5\pi / 6$ (${\alpha} = 5 \pi / 6$) [amongst other angles; not the same as the ones preferred by this magnetic field]. Thus a simple analytical solution for the scaling behaviour of the relative length change is not as easy to derive (nor illuminating). Nevertheless, just as for $\vec{H} \  || \  [100]$, the quadrupolar order parameters are even-functions-in-$h$ as can be seen in SI \ref{app_order_plot} Fig. \ref{110_magneto_all} and corroborated by numerical fits of the order parameters.

\subsection{$\vec{H} \parallel [111]$} \label{111_angle}

For magnetic fields applied along this direction, all the multipolar moments couple quadratically to the magnetic field. We first discuss the magnetic field dependency of the quadrupolar moments. For $T> T_{\mathcal{Q}}$, both quadrupolar moments are zero [as depicted in the numerical solution Fig. \ref{111_magneto_all}(a) in SI \ref{app_order_plot}] due to AF$\mathcal{Q}$ not spontaneously ordering, and the lack of a linear-in-$|\phi|$ coupling to the magnetic field for the F$\mathcal{Q}$ moment. For $T < T_{\mathcal{Q}}$, the AF$\mathcal{Q}$ has spontaneously ordered, and hence also permits a finite (but small) parasitic F$\mathcal{Q}$ moment, with a weak quadratic-in-$h$ scaling for F$\mathcal{Q}$; thus $| \phi | \approx \big|_{T_{\mathcal{O}} < T < T_{\mathcal{Q}}}  \phi_0  +  \phi_h  h^2 $ here, as seen in Fig. \ref{111_magneto_all}(b). 

The quadratic-in-$h$ scaling of the octupolar moment can be observed by performing a simple analytical approximation, where for the sake of simplicity, we assume only pure F${\mathcal{O}}$ ordering (setting the quadrupolar moments to zero). This simplified Landau free energy is of the form,
\begin{equation}
\begin{aligned}
F [m] =  \left[ \frac{t_{m}}{2} - \frac{g_{\mathcal{O}}^2}{2c_{44}} (h^2) \right] m ^2 + u_m m^4 \ , 
\end{aligned}
\end{equation}
where we have taken $h_x = h_y = h_z = h / \sqrt{3}$. Extremizing this free energy, and assuming $T$ is far enough below the octupolar critical temperature such that $ \frac{g_{\mathcal{O}}^2 h^2 / c_{44}}{|t_m|} $ is small enough to Taylor-expand to yield
\begin{equation}
\begin{aligned}
m \approx \sqrt{\frac{ |t_m| }{4 u_m} } \left( 1 + \frac{g_{\mathcal{O}}^2 h^2}{2 c_{44} | t_m | } \right) \ .
\end{aligned}
\end{equation}
Thus substituting this (simple) approximate solution into Eq. \ref{eq_length_subbed_multipolar} yields a linear in $h$ behaviour in the length change (superimposed on a quadratic background). 
In the presence of all order parameters, we obtain (from thorough numerical minimization of the full Landau free energy) the same scaling relation of the octupolar (and quadrupolar) moments i.e. $m \sim a_0 + a_2 h^2$, where $a_{0,2}$ are finite constants. 
The even-in-$h$ scaling behaviour of the multipolar moments is apparent qualitatively in SI \ref{app_order_plot} Fig. \ref{111_magneto_all}(c).
It is important to note that this scaling behaviour was performed under the assumption that we could neglect the $\mathcal{O}(h^3)$ coupling between the octupolar moment and the magnetic field i.e. neglected $h_x h_y h_z m$ in the Landau free energy. For magnetic fields along [100] and [110] this term is zero, and plays no role anyway, but for magnetic fields along the [111] direction it is non-zero (albeit small relative to the $\sim h^2 m^2$ coupling). More importantly, the cubic-in-$h$ coupling breaks the $\mathbb{Z}_2$ symmetry ($m \leftrightarrow -m$) of the octupolar moment. This introduces a `flip' in the octupolar moment at $h=0$ (and at $T < T_{\mathcal{O}}$ where $m$ has spontaneously ordered i.e. $m \neq0$):
for $h = 0^{+}$, the $+ |m|$ solution is `chosen', and as we crossover to $h=0^{-}$, the now physically distinct $- |m|$ solution is `chosen' (this is seen in Fig. \ref{fig:m_hysteresis_plot} in the next section). A similar phenomena is observed in usual ferromagnetism, below the ordering temperature.
Although this $\sim h^3 m$ correction can be safely ignored for discussions regarding scaling behaviour at small magnetic fields, its effect becomes very important when studying hysteresis behaviour.

\section{Hysteretic Behaviour of Octupolar Ordering} \label{sec_hyst_octupolar}
We are motivated in this section by recent unpublished experiments \cite{sakai_mpipks_2018} where {\it hysteretic} behaviour is observed in the length change along the [111] 
direction below the supposed-octupolar temperature.
Hysteresis arises from the existence of domains and the motion of domain walls in the presence of obstructing `pinning sites', which have not been taken into account in the
Landau theory we have studied. In order to incorporate such effects, we adapt the phenomenological approach due to Jiles and Atherton 
\cite{jiles_atherton_hysteresis, smith_hysteresis} which has been used to study hysteresis loops in ferromagnetic and ferroelastic materials. This approach identifies
the order parameter (obtained by minimizing the Landau free energy) as its ideal bulk value, 
where the Landau theory includes a direct coupling $u_f m h^3$ of the ferro-octupolar moment $m$ and the external [111] magnetic
field. Deviations from this ideal value are captured in terms of a `lag' $(m-m_{\rm ir})$, described by
a phenomenological equation
 \begin{equation} \label{equ:diff_ir}
 \frac{d m_{\rm ir} }{dh} = \frac{ m - m_{\rm ir} }{\pm k  - \alpha (m - m_{\rm ir}) }   (3  h^2)  \ ,
 \end{equation}
where $k$ characterizes the pinning strength (encoding the number of pinning sites and the energy cost of overcoming a single pinning site), $\alpha$ is a 
constant describing the coupling between octupolar domains, and the sign $\pm$ applies respectively for increasing and decreasing magnetic fields.
The experimentally relevant octupole moment is then given by
\begin{align} 
m_{\rm exp} = {m}_{\rm ir} +  c (m- {m}_{\rm ir}) \ .
\label{equ:total}
\end{align}
where $c$ is a constant. A heuristic derivation of these equations is given in the SI \ref{app_oct_domain}.
The algorithm to determine the total macroscopic octupolar moment is straightforward. First we minimize the Landau free energy to obtain the ideal octupolar moment
$m$. Next, we solve Eq. \ref{equ:diff_ir} for $m_{\rm ir}$. Finally, we obtain $m_{\rm exp}$ by using Eq. \ref{equ:total}.

\begin{figure}[t]
\centering
  \includegraphics[width=0.87\linewidth]{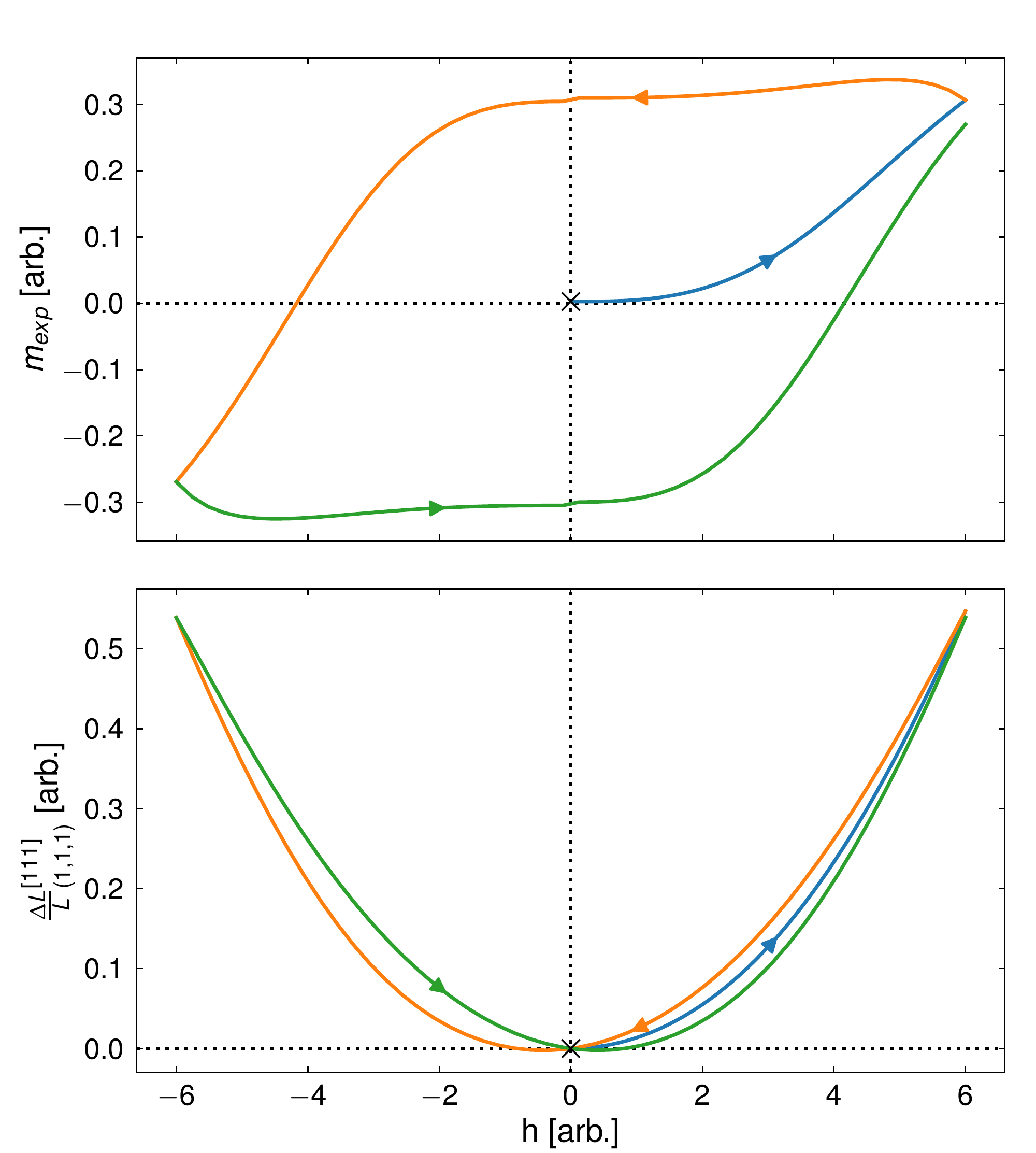} 
  \caption{Hysteresis for $\vec{H} \ || \ [111]$. (a) Total octupolar order parameter ($m_{\rm exp}$) versus magnetic field strength ($h$) along [111] direction demonstrating hysteresis for $T < T_{\mathcal{Q}}, T_{\mathcal{O}}$. Initial condition [denoted by `$\times$' in Figs. \ref{fig:m_hysteresis_plot}(a,b)] of $m_{\rm ir}=0$ for $h=0$ is used to obtain this solution, and $k=100$, $\alpha = 10^{-3}$, $c = 0.01$. (b) Length change along (1,1,1) direction demonstrating hysteresis, using the solution of \ref{fig:m_hysteresis_plot}(a), and taking $\gamma = 0.8$.}
      \label{fig:m_hysteresis_plot}
\end{figure}
 
We now examine hysteresis behaviour occurring at a temperature of $T< T_{\mathcal{O}}$.
Figure \ref{fig:m_hysteresis_plot}(a) depicts the hysteresis behaviour for the total bulk octupolar moment $m_{\rm exp}$, which is reminiscent of the hysteresis in ferromagnets. The initial condition used to solve Eq. \ref{equ:diff_ir} is chosen so that at $h=0$, the ideal configuration is not being met (i.e. $m_{\rm exp} \neq m$); 
this depicts the realistic scenario of having not \textit{all} domains aligned in the same direction at $h=0$. 
Inserting this solution for the octupolar moment into Eq. \ref{eq_length_subbed_multipolar}, we obtain the length change along the (1,1,1) direction, as shown in Fig. \ref{fig:m_hysteresis_plot}(b). For small magnetic field strengths, $\left(\frac{\Delta L }{L}\right)_{(1,1,1)}^{[111]} $ presents the linear in $h$ scaling as seen in Fig. \ref{fig:m_hysteresis_plot}(b).

\section{Conclusions} \label{conclusions_sec}

In this work, motivated by recent and ongoing experiments on Pr(Ti,V,Ir)$_2$(Al,Zn)$_{20}$, we have used Landau theory of multipolar orders coupled to lattice strain fields to
study magnetostriction in systems with quadrupolar and octupolar orders.
\textcolor{black}{Our theoretical results for magnetostriction in the presence of octupolar order 
appear consistent with recent magnetostriction experiments on PrV$_2$Al$_{20}$ where the onset of unusual linear-in-field and hysteretic magnetostriction is observed
for fields along the [111] direction for $T < 0.65$K \cite{sakai_mpipks_2018}. In addition, we can qualitatively understand the quadratic-in-field background magnetostriction observed in these experiments.}

In particular, we have discussed 
the scaling behaviour of the length change in such Landau theory models for magnetic fields applied along [100], [110] and [111] directions in 
three temperature regimes: $T>T_{\mathcal{Q}}, T_{\mathcal{O}}$, $T_{\mathcal{O}} < T < T_{\mathcal{Q}}$, and $T<T_{\mathcal{Q}}, T_{\mathcal{O}}$.
From our studies we conclude that linear-in-$h$ scaling of the length change is observed for length changes (both parallel and perpendicular) to magnetic fields applied along the [111] direction, and for length changes perpendicular to [100], below $T_{\mathcal{O}}$. Moreover, the coefficient of the linear-in-$h$ term is directly proportional to the octupolar moment, thus giving a distinct signature for the onset of octupolar ordering as well as a means to detect/measure the octupolar moment. For magnetic fields applied along the [100] and [110] directions, the length changes also acquire quadratic-in-$h$ scaling behaviour. This scaling arises from the quadrupolar moments and/or direct coupling of the conduction electrons to the external magnetic field and the lattice normal modes. The summary of the scaling behaviours is presented succinctly in Table \ref{tab:summary_scaling_all}. Finally, we demonstrate a characteristic hysteresis behaviour in the octupolar order parameter and the associated (1,1,1) length change for magnetic field applied along [111]. This hysteretic
 behaviour is identified as a consequence of the $\sim h^3 m$ direct magnetic field coupling to the octupolar moment.

In terms of future work, an interesting avenue to explore is that of the coupling of the conduction electrons to the multipolar moments, as well as to the lattice strain and magnetic field. In particular, the origin of the conduction electron term in Eq. \ref{f_strain_oct}, introduced in our phenomenological model from symmetry arguments, is a fascinating direction to explore \textcolor{black}{(as well as potential other terms arising from conduction electrons)}. Understanding the nature and role of the conduction electrons will also help shed light on the quantum critical behaviour and superconductivity in such multipolar Kondo lattice systems \cite{hfm_superconductivity_v, pressure_hfm_super_quad_pr_ti, super_fq_pr_ti, cox_1, two_channel_emery, cox_2,u_quad_kondo, pr_nfl, pr_nfl_v_ti,quad_nfl_onimaru,Kusunose_nfl_2016,pr_v_nfl}.

\matmethods{
\textcolor{black}{The Landau theory is numerically minimized using standard optimization schemes. The hysteresis differential equation is numerically solved using Runge-Kutta 4$^\text{th}$ order methods.}}


\acknow{We thank Piers Coleman and Premi Chandra for helpful discussions on Landau theory of magnetostriction, in particular relating to the coupling between the octupole and magnetic field. We also thank Wonjune Choi and Li Ern Chern for helpful comments regarding the manuscript. This work was supported by NSERC of Canada, and Canadian Institute for Advanced Research. S.B.L. is supported by the KAIST startup and National Research Foundation Grant (NRF-2017R1A2B4008097). This work was partially supported by Grants-in-Aids for Scientific Research on Innovative Areas (15H05882 and 15H05883) from the Ministry of Education, Culture, Sports, Science, and Technology of Japan,  by CREST(JPMJCR18T3), Japan Science and Technology Agency, and by Grants-in-Aid for Scientific Research (16H02209) from the Japanese Society for the Promotion of Science (JSPS).
}
\showacknow 



\bibliography{magnetostriction_references_bibtex}

\leadauthor{Patri} 

\newpage
\clearpage

\beginsupplement
\section*{Supporting Information (SI)}

\subsection{Symmetry transformations of multipolar order parameters} \label{app_multipolar_symm}

Under the symmetry constraints detailed in the main text Sec. \ref{sec_landau}, the multipolar moments transform as denoted in Table \ref{tab_order}.

\setlength{\extrarowheight}{2pt}
\begin{table}[h]
\begin{tabular}{c|c|c|c|c}
\toprule
\hline
Symmetry & F$\mathcal{Q}$ & F$\mathcal{O}$ & AF$\mathcal{Q}$ & AF$\mathcal{O}$  \\
\hline 
$\mathcal{I} $& $\phi \rightarrow \phi$ & $m \rightarrow m$ & $\tilde{\phi} \rightarrow -\tilde{\phi}$ & $\tilde{m} \rightarrow -\tilde{m}$ \\
${\Theta} $& $\phi \rightarrow \phi^*$ & $m \rightarrow -m$ & $\tilde{\phi} \rightarrow \tilde{\phi}^*$ & $\tilde{m} \rightarrow -\tilde{m}$ \\
$\mathcal{S}_{4z} $& $\phi \rightarrow -\phi^*$ & $m \rightarrow -m$ & $\tilde{\phi} \rightarrow -\tilde{\phi}^*$ & $\tilde{m} \rightarrow -\tilde{m}$ \\
$\mathcal{C}_{31} $& $\phi \rightarrow e^{-i \frac{2 \pi}{3}} \phi$ & $m \rightarrow m$ & $\tilde{\phi} \rightarrow e^{-i \frac{2 \pi}{3}} \tilde{\phi}$ & $\tilde{m} \rightarrow \tilde{m}$ \\
\hline
\bottomrule
\end{tabular}
\caption{Transformation of multipolar order parameters under generating elements of $T_d$ ($\mathcal{S}_{4z}, \ \mathcal{C}_{31}$), bond centre inversion ($\mathcal{I}$) and time reversal ($\Theta$). The `$*$' indicates complex conjugation.}
\label{tab_order}
\end{table}
where in real $\mathbb{R}^3$ space, the matrix representations of $\mathcal{S}_{4z}$ and $\mathcal{C}_{31}$ are,

\noindent\begin{minipage}{.5\linewidth}
\[
  \mathcal{S}_{4z} = 
    \begin{bmatrix}
0 & -1 & 0 \\
1 & 0 & 0 \\
0 & 0 &1\\
  \end{bmatrix}
  \cdot \mathbb{I},
\] \\
\end{minipage}%
\begin{minipage}{.5\linewidth}
\[
   \mathcal{C}_{31} = 
    \begin{bmatrix}
0 & 0 & 1 \\
1 & 0 & 0 \\
0 & 1 & 0 \\
  \end{bmatrix}, 
\] \\
\end{minipage}
where $\mathbb{I}$ denotes parity $(x,y,z) \rightarrow (-x, -y, -z)$.

\subsection{General expression of length change in different directions} \label{app_length_derivation}

The relative length change, $\Delta L / L$, of the crystal can be shown to be related to the components of the strain tensor. To derive such an expression we consider a pair of neighbouring points ($A$ and $B$) situated at $\vec{x}$ and $\vec{x} + \Delta \vec{x}$, respectively, in the undeformed crystal. The vector separating these two points is $\Delta \vec{x} = \Delta s \hat{\ell}$, where $\Delta s$ is the distance between A and B, and $\hat{\ell} \equiv \hat{e}_{A \rightarrow B}$ is the unit vector directed from $A$ to $B$. Under a deformation, the points $A$ and $B$ get respectively shifted by displacement vectors $\vec{u}$ and $\vec{u}(\vec{x} + \Delta \vec{x})$ to new locations $\vec{y}(\vec{x}) =  \vec{x} + \vec{u} (\vec{x})$ and $\vec{y}(\vec{x}+\Delta\vec{x}) = [\vec{x} + \Delta \vec{x}] + \vec{u}(\vec{x} + \Delta \vec{x})$. The relative vector connecting these neighbouring points in the deformed crystal is $\Delta \vec{y} = \Delta \vec{x} + \Delta \vec{u}$, where for small enough displacements $\Delta u_i = \Sigma_{k} \frac{\partial u_i}{\partial x_k} \Delta x_k$; $i = 1,2,3$ is the unit vector directions, and $k$ is being summed over. Expanding the relative length change ($|\Delta \vec{y}| - |\Delta \vec{x}|$) over the initial separation of the points ($|\Delta \vec{x}|$) leads to the general expression of the length change along a direction $\vec{\ell}$

\begin{equation}
\left(\frac{\Delta L }{L}\right)_{\vec{\ell}} = \sum_{i,j=1} ^{3} \epsilon_{ij} \hat{\ell}_i \hat{\ell}_j \ , 
\end{equation}

where $\epsilon_{ij} \equiv  \frac{1}{2}\left (\frac{\partial u_i}{\partial x_j} + \frac{\partial u_j}{\partial x_i} \right) $ is the familiar strain tensor, and $\hat{\ell}_i$ is the $i^{th}$ component of the unit vector $\hat{\ell}$.
\subsection{Values of Landau Parameters} \label{app_landau_params}

The values of the Landau parameters are arbitrary to a certain extent, and depending on the choice of the parameters the subsequent scaling coefficients are altered. For the studies in this work, we used the following values for the Landau parameters (in the appropriate units): $T_{\mathcal{Q}}^c = 40$ and $T_{\mathcal{O}}^c = 6$, $u_{\tilde{\phi}} = u_{{m}} = u_{{\phi}} = w_{\tilde{\phi}} = -u_{\tilde{\phi}m} = - u_{{\phi}m} = 5$, $l_{\tilde{\phi}} = 6$, $v_{{\phi}} = -2$, $\tilde{r}_H = r_H = 0.05$, $\tilde{s}_H = s_H = -0.0006$, $(g_1, g_2) = (-1.7, 3.7)$, $g_{\mathcal{O}} = 2.19$, $g_{\mathcal{Q}} = 0.4$, $(c_{11} - c_{12}) = 10^2$, and $ c_{44} = 2.4 \times 10^2$. The values for the elastic modulus tensor components are chosen to be large, because as seen in  Eq. \ref{landau_free_all} they are only responsible for shifting the critical temperatures (the mass term) of AF$\mathcal{Q}$, F$\mathcal{O}$ (F$\mathcal{Q}$).

\subsection{Multipolar Order parameters with respect to magnetic field strength}  \label{app_order_plot}

\begin{figure*} [t]
\centering
  \includegraphics[width=0.9\linewidth]{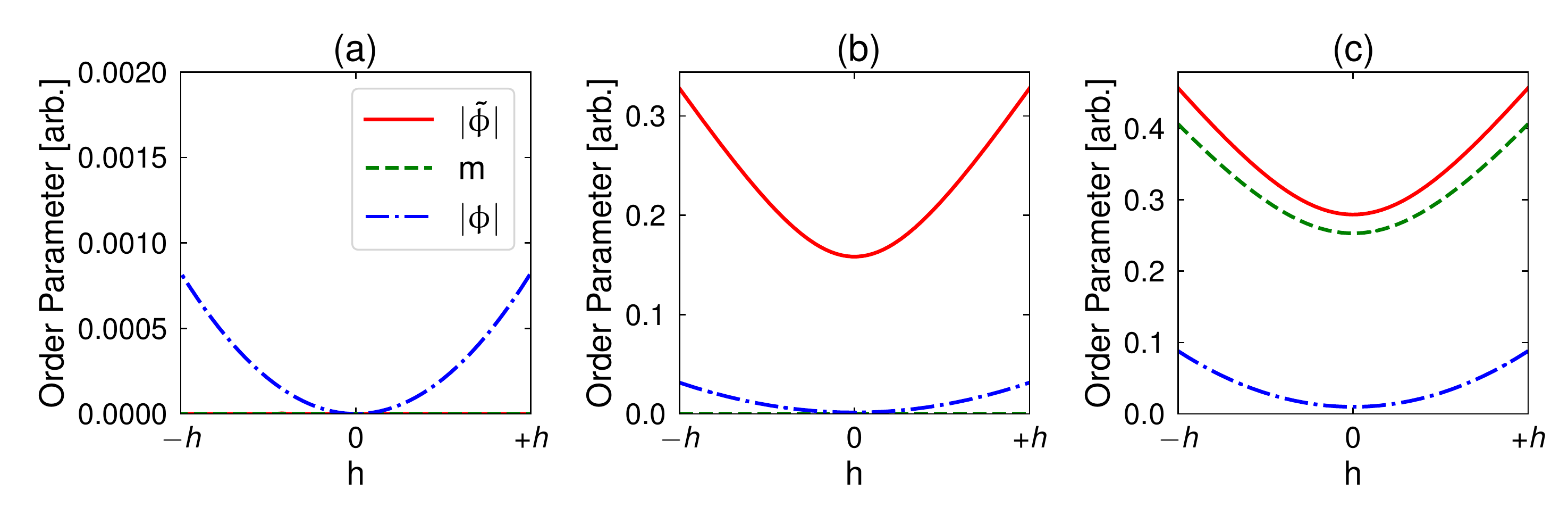}  
  \caption{Order parameters AF$\mathcal{Q}$ [$|\tilde{\phi}|$], F$\mathcal{O}$ [$m$], F$\mathcal{Q}$ [$|{\phi}|$] versus magnetic field strength $h$ applied along [100] direction, for (a) $T > T_{\mathcal{Q}}, T_{\mathcal{O}}$ , (b) $T_{\mathcal{O}} < T < T_{\mathcal{Q}}$, and (c) $T < T_{\mathcal{Q}}, T_{\mathcal{O}}$. Qualitatively, the multipolar moments possess even-in-$h$ symmetry.}
   \label{100_magneto_all}
\end{figure*}
\begin{figure*} [t]
\centering
  \includegraphics[width=0.9\linewidth]{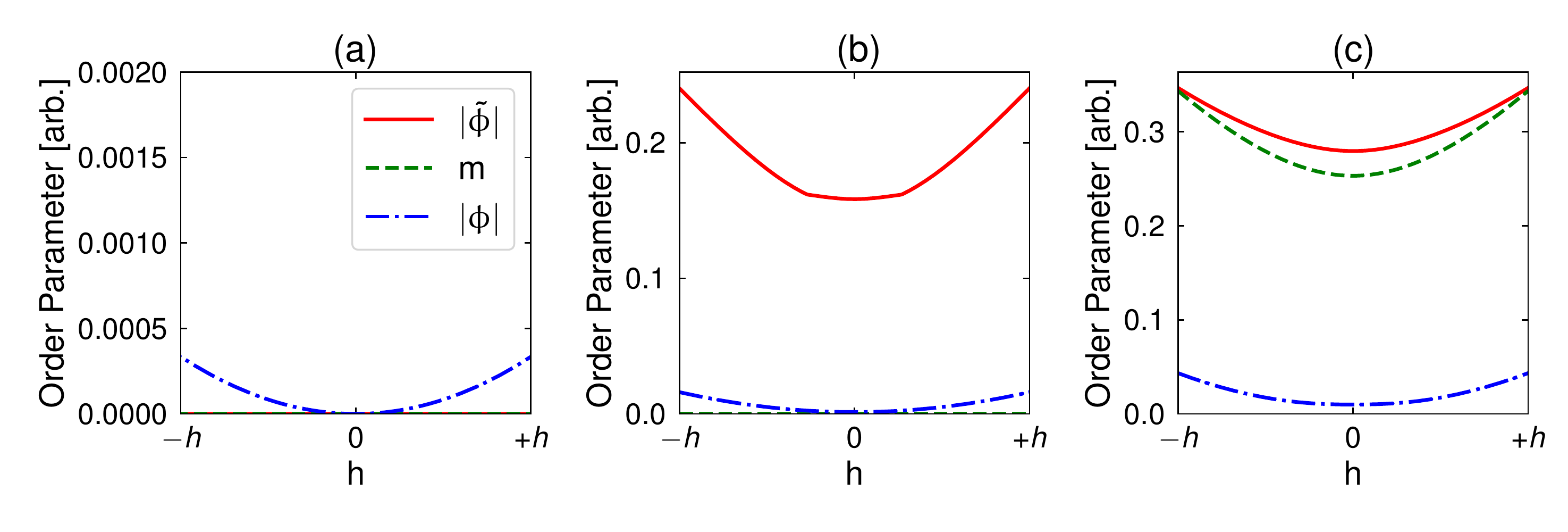}
  \caption{Order parameters AF$\mathcal{Q}$ [$|\tilde{\phi}|$], F$\mathcal{O}$ [$m$], F$\mathcal{Q}$ [$|{\phi}|$] versus magnetic field strength $h$ applied along [110] direction, for $T > T_{\mathcal{Q}}, T_{\mathcal{O}}$ , (b) $T_{\mathcal{O}} < T < T_{\mathcal{Q}}$, and (c) $T < T_{\mathcal{Q}}, T_{\mathcal{O}}$. Qualitatively, the multipolar moments possess even-in-$h$ symmetry.}
  \label{110_magneto_all}
\end{figure*}
\begin{figure*} [t!]
\centering
  \includegraphics[width=0.9\linewidth]{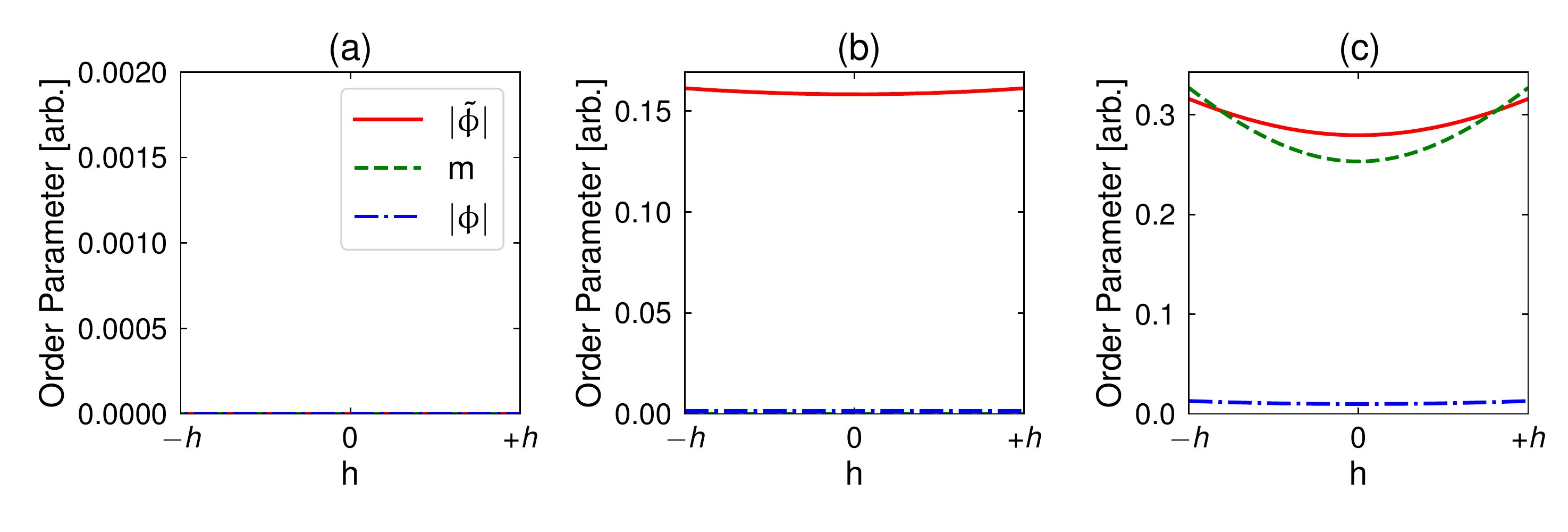}
  \caption{Order parameters AF$\mathcal{Q}$ [$|\tilde{\phi}|$], F$\mathcal{O}$ [$m$], F$\mathcal{Q}$ [$|{\phi}|$] versus magnetic field strength $h$ applied along [111] direction, for (a) $T > T_{\mathcal{Q}}, T_{\mathcal{O}}$ , (b) $T_{\mathcal{O}} < T < T_{\mathcal{Q}}$, and (c) $T < T_{\mathcal{Q}}, T_{\mathcal{O}}$. Qualitatively, the multipolar moments possess even-in-$h$ symmetry.}
  \label{111_magneto_all}
\end{figure*}

Figures \ref{100_magneto_all}, \ref{110_magneto_all}, \ref{111_magneto_all} presents the solutions of the order parameters from a thorough numerical study of the complete Landau free energy. As can be seen, the F$\mathcal{Q}$ moment is indeed an even function-in-$h$. 

We stress on the atypical nature of the $[110]$ direction, in that it introduces a degeneracy into the system. We can observe this by just focusing on the AF$\mathcal{Q}$ moment's terms, where if we take
\begin{align}
&\tilde{\alpha} = 0 + x && \implies \cos(6 \tilde{\alpha}) = \cos(6 x)    \ ,  \nonumber \\
&\tilde{\alpha} = 0 + x && \implies  \sin(-\pi / 2  +  2 \tilde{\alpha}) =  - \cos(2 x) \ ,   \nonumber  \\
&{\alpha} = \pi/2 + y && \implies \sin(3 \tilde{\alpha}) = -\cos(3 y)   \ ,    \\
&{\alpha} = \pi/2 + y && \implies  \cos(\alpha - (-\pi/2)) =  -\cos( y) \ ,   \nonumber 
\end{align}
where $x$ and $y$ are the deviations from $0$ and $\pi/2$, respectively. These terms are invariant under $x \rightarrow -x$ and $y \rightarrow -y$, resulting in two equivalent orientations for the quadrupolar moment. It is this symmetry that is responsible for degeneracy and the observed `multiple' solutions. The `multiple' solutions persist for small magnetic field strengths (where the competition with the anisotropic term is still ongoing); however, for large enough field strengths, the magnetic field term dominates over the anisotropic term to yield one unique solution determined by the magnetic field term. The point at which the magnetic field has dominated and takes over the anisotropic term is observed as a `kink' in the order parameters, as seen in Fig. \ref{110_magneto_all}; above this $h_{kink}$ the `multiple' solutions vanishes, and a single/unique solution emerges. Note that this `multiple' solutions phenomena is not unique to the [110] field direction; it can readily be transferred over to the [100] direction's solutions by flipping the sign of the chosen parameters $v_{\phi}, \tilde{r}_g, g_1$, which then gives the pure/unique solution to [110]. Nevertheless, this degeneracy does not impact the observed scaling (even-in-$h$) behaviour.

In temperature regimes of $T<T_{\mathcal{Q}}$, there is the aforementioned kink at $h_{kink}$ in both the AF$\mathcal{Q}$ and F$\mathcal{Q}$ order parameters for $T_{\mathcal{O}} <T < T_{\mathcal{Q}}$ and $T<T_{\mathcal{Q}}, T_{\mathcal{O}}$, respectively, as seen in Fig. \ref{110_magneto_all}(b,c); the kink is more noticeable in the $|\tilde{\phi}|$ than $|\phi|$ due to the small, parasitic nature of F$\mathcal{Q}$.  It is interesting to note that this `kink' appears for a larger value of $h$ for $T<T_{\mathcal{Q}}, T_{\mathcal{O}}$ than in $T_{\mathcal{O}} <T < T_{\mathcal{Q}}$. This stems from the fact that for $T$ much lower than $T_{\mathcal{Q}}$, the $|\tilde{\phi}|^6$ and $|{\phi}|^3$ contribution to the free energy is substantial, and thus to overcome these anisotropic terms it requires a larger magnetic field. Hence, the value of $h_{kink}$ grows the lower in $T$ we go (below $T_{\mathcal{Q}}$). Apart from this `kink', the scaling relations of the F$\mathcal{Q}$ is as expected with a constant zero-in-$h$ shift being present for $T<T_{\mathcal{Q}}$, due to presence of spontaneously ordered AF$\mathcal{Q}$. We also note that for large enough $h$, the AF$\mathcal{Q}$ is tuned to zero with the ferro-like moments surviving.

\subsection{Domain Wall model of octupolar moments} \label{app_oct_domain}

We provide a more detailed derivation of the hysteresis model of octupolar moments. The derivation follows that of Jiles and Atherton \cite{jiles_atherton_hysteresis} with a few modifications that we elaborate on. 

The basic premise is that of two equally large (in terms of volume) domains of octupolar order separated by a domain wall. This domain wall is assumed to lie directly on top of a so-called pinning site. A pinning site can be any object that obstructs the motion of domain walls under the influence of a magnetic field; the true nature of the pinning site is not of great importance to the derivation presented here. The domains possess an octupolar moment per unit volume, $m_d$.
\textcolor{black}{Due to the Ising-like nature of the octupolar moment, one can think of one domain being $m_d = +m_d$, while the other domain being $m_d' = -m_d$. The interaction of each domain's octupolar moment with the magnetic field and the bulk octupolar moment of the system is described by $E_{coupling} = -u_f  m_d (h_x h_y h_z + \alpha m) $, where $u_f$ is a coefficient of coupling, and where we incorporate inter-domain coupling by a Weiss-like mean field term ($\alpha$). $f_e \equiv h_x h_y h_z + \alpha m$ is the effective field.}

We now consider the application of a magnetic field on the system that encourages the expansion of the $+m_d$ domain i.e. it is energetically favourable to have both the domains align as $+m_d$. In the absence of the pinning site, this domain wall slides over easily thus enabling the expansion of the domain. However, the pinning site obstructs this simple motion. In the spirit of Jiles and Atherton, we consider the energy required to overcome the pinning site to be equal (up to a proportionality constant, $c_0$) to the energy required to align the octupolar moment of $-m_d$ with $+m_d$ i.e.
\begin{equation}
\begin{aligned}
E_{cost} &= c_0 \left( u_f m_d f_e - (u_f m_d' f_e) \right) \\
		& = c_0 \left(2 u_f m_d f_e \right) \ .
\end{aligned}
\end{equation}
This is the energy required to overcome a single pinning site. We now generalize the scenario where we have a collection of such pinning sites over a distance $dx$, and the magnetic field is applied such that the domain wall (of cross sectional area, $A$) is swept over that distance $dx$. If there exists an average density of pinning sites, $n$, over this volume $A \ dx$ and the average energy to overcome a site is $\langle E_{cost} \rangle$ [where the averaging is performed over all the pinning sites] then the total energy dissipated through moving this domain wall through this mire of pinning sites is
\begin{equation}
\begin{aligned}
E_{total}(x) &= \int_{0}^{x} \langle E_{cost} \rangle n A dx  \ , 
\end{aligned}
\end{equation}
where $n A dx$ is the total number of pinning sites in the volume of interest. Since the change in the bulk octupolar moment (as the magnetic field moves the domain wall past the pinning site) is $d m_{\rm {ir}} = \left(m_d - (-m_d) \right)A dx = 2 m_d Adx$, then we can replace the integrand in the above equation by
\begin{equation}
\begin{aligned}
E_{total}(x) &= \int_{0}^{{m}_{ir}} \frac{n \langle E_{cost} \rangle}{2 {{m}_{d}}} d {{m}_{\rm {ir}}} \\
		& = k \int_{0}^{{m}_{ir}}  d {{m}_{\rm {ir}}} \ , 
\label{app_oct_cost}		
\end{aligned}
\end{equation}
where $k \equiv \frac{n \langle E_{cost} \rangle}{2 m_{d}}$, which we take to be a constant. Although the derivation is quite involved, the final result physical makes sense in that the total work done in moving the domain wall past the pinning sites is proportional to the change in the bulk octupolar moment (associated with moving the domain wall).

Next, we consider the energy required to set up a bulk octupolar configuration, ${m}_{\rm {ir}}$ (under the presence of an effective field, $f_e$) which is the energy required to setup the ideal bulk octupolar moment in the absence of pinning sites, $m$, in an effective field, $f_e$, $\textit{plus}$ the energy required to overcome the domain wall,
\begin{equation}
\begin{aligned}
- \int {m}_{\rm {ir}} d f_e &= -\int {m} d f_e + k \int d {m}_{\rm {ir}}  \\
				& = -\int {m} d f_e + k \int \frac{d {m}_{\rm {ir}}}{d f_e} d f_e
\end{aligned}
\end{equation}
Collecting the integrands,
\begin{equation}
\int \left( {m}_{\rm {ir}} - {m} + k \frac{d {m}_{\rm {ir}}}{d \tilde{f_e}} \right) d \tilde{f_e} =0 \ , 
\label{eq:diff_integrand}
\end{equation}
where we introduce the dummy variable of integration $\tilde{f_e}$. For magnetic fields being swept up from 0 to $f_e$, the bounds on the integrand are: $\int_{up} \equiv \int_{0} ^{f_e}$, while for sweeping down from 0 to $-{f_e}$ is from $\int_{down} \equiv \int_{0} ^{-f_e}$. Taking the derivative of the above Eq. \ref{eq:diff_integrand} for the up sweep we get (by applying the Fundamental theorem of Calculus),
\begin{equation}
\begin{aligned}
& \frac{d}{d f_e} \left( \int_{0} ^{f_e} \left( {m}_{\rm {ir}} - {m} + k \frac{d {m}_{\rm {ir}}}{d \tilde{f_e}} \right) d \tilde{f_e} =0 \right) \\
& \implies  {m}_{\rm {ir}} = {m} - k \frac{d {m}_{\rm {ir}}}{d {f_e}} \ . 
\end{aligned}
\end{equation}
Similarly for the down sweeping magnetic field, we get (by changing variables in the integrand $\tilde{f_e} \rightarrow - \tilde{f_e}$),
\begin{equation}
\begin{aligned}
& \frac{d}{d f_e} \left( \int_{0} ^{-f_e} \left( {m}_{\rm {ir}} - m + k \frac{d {m}_{\rm {ir}}}{d \tilde{f_e}} \right) d \tilde{f_e} =0 \right) \\
& \implies  {m}_{\rm {ir}} = {m} + k \frac{d {m}_{\rm {ir}}}{d {f_e}}\ . 
\end{aligned}
\end{equation}
Thus, we obtain ${m}_{\rm {ir}} = {m} - \pm k \frac{d {m}_{\rm {ir}}}{d {f_e}}$, where $\pm$ is for increasing and decreasing magnetic fields, respectively. Rearranging, we arrive at the following differential equation
\begin{equation}
 \frac{d {m}_{\rm {ir}}}{df} = \frac{ {m} - {m}_{\rm {ir}} }{\pm k  - \alpha ({m} - {m}_{\rm {ir}}) }    \ .
 \label{eq:hyst_appendix}
 \end{equation}
Finally, we use$\frac{d {m}_{\rm {ir}}}{df} = \frac{d {m}_{\rm {ir}}}{dh} \frac{dh}{df}$ to obtain Eq. \ref{equ:diff_ir} in the main text.

\end{document}